\documentclass[a4paper]{cas-dc} 

\usepackage[numbers]{natbib}

\usepackage{relsize}
\usepackage{graphicx}
\usepackage{subcaption}
\usepackage[boxruled,vlined]{algorithm2e}
\usepackage{amsmath}
\usepackage{xcolor}
\usepackage{float}
\usepackage{soul}
\usepackage{etoolbox}
\usepackage{setspace}
\usepackage{gensymb}
\usepackage{placeins}
\usepackage[font={small,sf}]{caption}
\def\tsc#1{\csdef{#1}{\textsc{\lowercase{#1}}\xspace}}
\tsc{WGM}
\tsc{QE}
\tsc{EP}
\tsc{PMS}
\tsc{BEC}
\tsc{DE}

\begin{document}
\let\WriteBookmarks\relax
\def\floatpagepagefraction{1}
\def\textpagefraction{.001}
\let\printorcid\relax
\shorttitle{Identification and Tracking of Deformation Twin Structures}

\shortauthors{H. J. Ehrich et al.}

\title [mode = title]{Automated Identification and Tracking of Deformation Twin Structures in Molecular Dynamics Simulations}





\author[1]{H. J. Ehrich}
\credit{Methodology, Software, Visualization, Writing - Original Draft, Data Curation}

\author[2,3]{A. Dollmann}
\credit{Validation, Writing - Review \& Editing}

\author[1]{P. G. Grützmacher}
\credit{Validation, Writing - Review \& Editing}

\author[1]{C. Gachot}
\credit{Supervision, Writing - Review \& Editing}


\author[1,4]{S. J. Eder}
                        %
\cormark[1]
\ead{stefan.j.eder@tuwien.ac.at}
\credit{Conceptualization, Supervision, Writing - Review \& Editing, Project administration, Data Curation}


\affiliation[1]{organization={Institute for Engineering Design and Product Development, TU Wien},
    addressline={Lehárgasse 6 – Objekt 7}, 
    city={1060 Vienna},
    country={Austria}}
\affiliation[2]{organization={Institute for Applied Materials (IAM), Karlsruhe Institute of Technology (KIT)},
    addressline={Kaiserstraße 12}, 
    city={76131 Karlsruhe}, 
    country={Germany}}
\affiliation[3]{organization={MicroTribology Center (µTC)},
    addressline={Straße am Forum 5},
    city={76131 Karlsruhe}, 
    country={Germany}}
\affiliation[4]{organization={AC2T research GmbH},
     addressline={ Viktor-Kaplan-Straße 2/C}, 
    city={2700 Wiener Neustadt},
    country={Austria}} 

\cortext[cor1]{Corresponding author}
\begin{abstract}
Deformation twinning significantly influences the microstructure, texture, and mechanical properties of metals, necessitating comprehensive studies of twin formation and interactions. 
While experimental methods excel at analyzing individual samples, they often lack the capability for temporal analysis of twinned structures.
Molecular dynamics simulations offer a temporal dimension, yet the absence of suitable tools for automated crystal twin identification has been a significant limitation.
In this article, we introduce a novel computational tool integrated into the visualization and analysis software OVITO.
Our tool automates the identification of coherent twin boundaries, links related twin boundaries, validates twin structures through orientation analysis, and tracks twins over time, providing quantifiable data and enabling in-depth investigations.
Validation on a copper single crystal under shear loading demonstrates successful tracking of various twins, revealing their genesis and growth over multiple timesteps. 
This innovative approach promises to advance the computational materials science domain by facilitating the study of deformation twinning, offering profound insights into the behavior and mechanical performance of materials.
\end{abstract}


\maketitle

\section{Introduction}

In the realm of materials science and engineering, the mechanical properties of metals and alloys play a pivotal role in determining the structural stability, wear resistance, and longevity of components~\cite{zhai2021recent}. 
Responsible for the mechanical properties are the underlying deformation mechanisms, with deformation twinning being one of the prominent avenues for plastic deformation~\cite{venables1961deformation, mahajan1973deformation, christian1995deformation}.
Deformation twinning involves the homogeneous shear of atomic layers within crystal lattices, resulting in the creation of distinct twin boundaries that separate regions of mirror-image symmetry.
While slip is the predominant deformation mechanism for metals, twinning constitutes the second of the two primary mechanisms through which metals (particularly those with few independent slip systems) accommodate external stresses, thereby shaping their behavior under various loading conditions~\cite{meyers2001onset}.

Twinning can be significantly pronounced in materials with low stacking fault energy~\cite{martin2016deformation,kwon2010textureevolution}, at high strain rates~\cite{grassel2000high,Frommeyer2003supra,christian1995deformation} or at low temperatures~\cite{kwon2010textureevolution,blewitt1957low,mori1977twinning,ALLAIN2004158}. 
These conditions promote an increase in stacking fault width and therefore the occurrence of deformation twinning.
Deformation twinning, in turn, enhances a material's strength and ductility~\cite{coomanStateofknowledge} by increasing the work hardening behavior~\cite{martin2016deformation,GUTIERREZURRUTIA20116449,KARAMAN20001345,IDRISSI2010961}. 
Deformation twins act as obstacles for dislocation glide, reducing the mean free path of dislocations~\cite{CHOI201827}, a phenomenon often referred to as the ``dynamic Hall-Petch effect"~\cite{GUTIERREZURRUTIA20116449,Barbier2009Analysis,BOUAZIZ2008484}.
In the context of steels, in the ductility-tensile strength or global formability diagram, twinning-induced plasticity (TWIP) steels lie above the range classically referred to as ``material banana"~\cite{WorldAutoSteel}.
In contrast to conventional steels, which often see a decrease in ductility as they become harder, TWIP steels, instead of losing ductility, actually exhibit an increase in the same~\cite{REMY197799}.

Understanding and characterizing deformation twinning have emerged as vital endeavors for both fundamental research and practical applications~\cite{yu2012nanostructured}.
The prevalence and evolution of twin boundaries dictate the material's response to stress, influencing aspects ranging from work hardening to crack initiation and propagation~\cite{greer2011plasticity}.

With the exception of some brave attempts, traditional experimental methods are often limited in their ability to capture the dynamic temporal evolution of twin structures during deformation processes~\cite{dollmann2022dislocation, dollmann2023temporal}.
These methods usually provide static snapshots of twinning events in isolated samples but struggle to reveal the intricate details of twin growth and interaction over time.
To bridge this gap and gain a comprehensive understanding of twinning mechanisms, computational tools have become indispensable.
Molecular dynamics (MD) simulations present a powerful avenue for investigating the dynamic evolution of materials at atomic scales~\cite{Argibay2017, eder2021elucidating, EDER2022101588}. 
These simulations offer the unique advantage of observing the temporal progression of materials under external stresses~\cite{grutzmacher2020visualization, eder2018interfacial}. 
However, despite their potential, MD simulations have long lacked a suitable tool to systematically identify and analyze twins automatically or at least semi-automatically. 
Such a tool would enable researchers to track twin formation, evolution, and interactions, facilitating a deeper understanding of twinning as a deformation mechanism.

This paper introduces a novel algorithm, built upon the OVITO platform~\cite{ovito}, that addresses this crucial gap in the field. 
The algorithm serves a multifaceted purpose, automating the identification of coherent twin boundaries, establishing connections between related twin boundaries, and ultimately tracking the temporal evolution of the identified twin structures. 
By seamlessly integrating these functionalities, the algorithm provides quantifiable data with an option of detailed investigations into individual twins.
To demonstrate the algorithm's efficacy, we applied it to study the deformation behavior of a copper single crystal subjected to shear. 
The algorithm successfully tracked and analyzed various twin boundaries, revealing insights into their genesis and growth over multiple timesteps.
Furthermore, the orientation analysis reliably validated twins, showcasing the algorithm's accuracy and utility in elucidating the intricate mechanisms that underlie twinning.

This paper addresses the critical need for a powerful tool to study deformation twinning through MD simulations. 
By introducing an innovative algorithm capable of identifying and tracking twins, this study empowers researchers with a robust platform to explore the fundamental aspects of twinning, fostering deeper insights into material behavior and its implications for mechanical performance.

\section{Used Software}
\label{sec:MandM}

The basis for any visualization and analysis of MD simulations is the per-atom data of a successfully conducted simulation run.
This atom or geometry data contains a snapshot of information such as position, velocity, stresses, order parameters, or other relevant properties of each atom in the simulated system.
The obtained data is then usually displayed and evaluated with a software different from the one that produced the raw data.


The required model systems should be characterized by homogeneous composition and microstructure (single-crystal) and only subjected to uniaxial load.
Based on established theories~\cite{KARAMAN20001345}, it is possible to predict their response to the ruling conditions and subsequently compare these predictions to the obtained results.
These simple custom samples were constructed to systematically test and validate the featured code.
AtomSK~\cite{HIREL2015212} was used to generate realistic single-crystalline face-centered cubic (fcc) specimens as an input for the MD simulation.
Simple normal-load cases were applied to single-crystalline structures using LAMMPS~\cite{LAMMPS}, an open-source molecular dynamics simulation software, in order to analyze unambiguous stress fields.
Shear simulations were then conducted for the final goal of being able to successfully analyze more complex stress fields.

The algorithm developed in the course of this work is based on the powerful visualization and analysis software OVITO~\cite{ovito}.
Important functions include the reliable recognition of crystal structures and their spatial position and rotation, determination of property relationships between neighboring atoms, and tracking of atomic motion over multiple time steps.
Unfortunately, the predefined so-called modifiers in OVITO do not allow the user to identify twinned structures by default.
However, the possibility to include custom Python scripts expands OVITO's capabilities considerably.
As the stacking of the atoms is reversed at the twin boundary, twin boundaries in fcc crystals appear as single layers of atoms with hexagonal-close-packed (hcp) structure in OVITO.
In contrast, stacking faults are detected as two adjacent layers of hcp-like atoms~\cite{ovitotwindef}, which form the next-most similar structure.
This implies a difference between both patterns concerning an atom's number of nearest neighbors with the same structure type, which can be exploited to find planar structures.
One of the important features of the presented tool is to robustly disambiguate between stacking faults and proper twins.
Parameterized from the acquired data, planes can be compared to find crystallographically meaningful connections.

\section{Implementation}

\begin{figure*}[t]
    \centering
    \includegraphics[width=\textwidth]{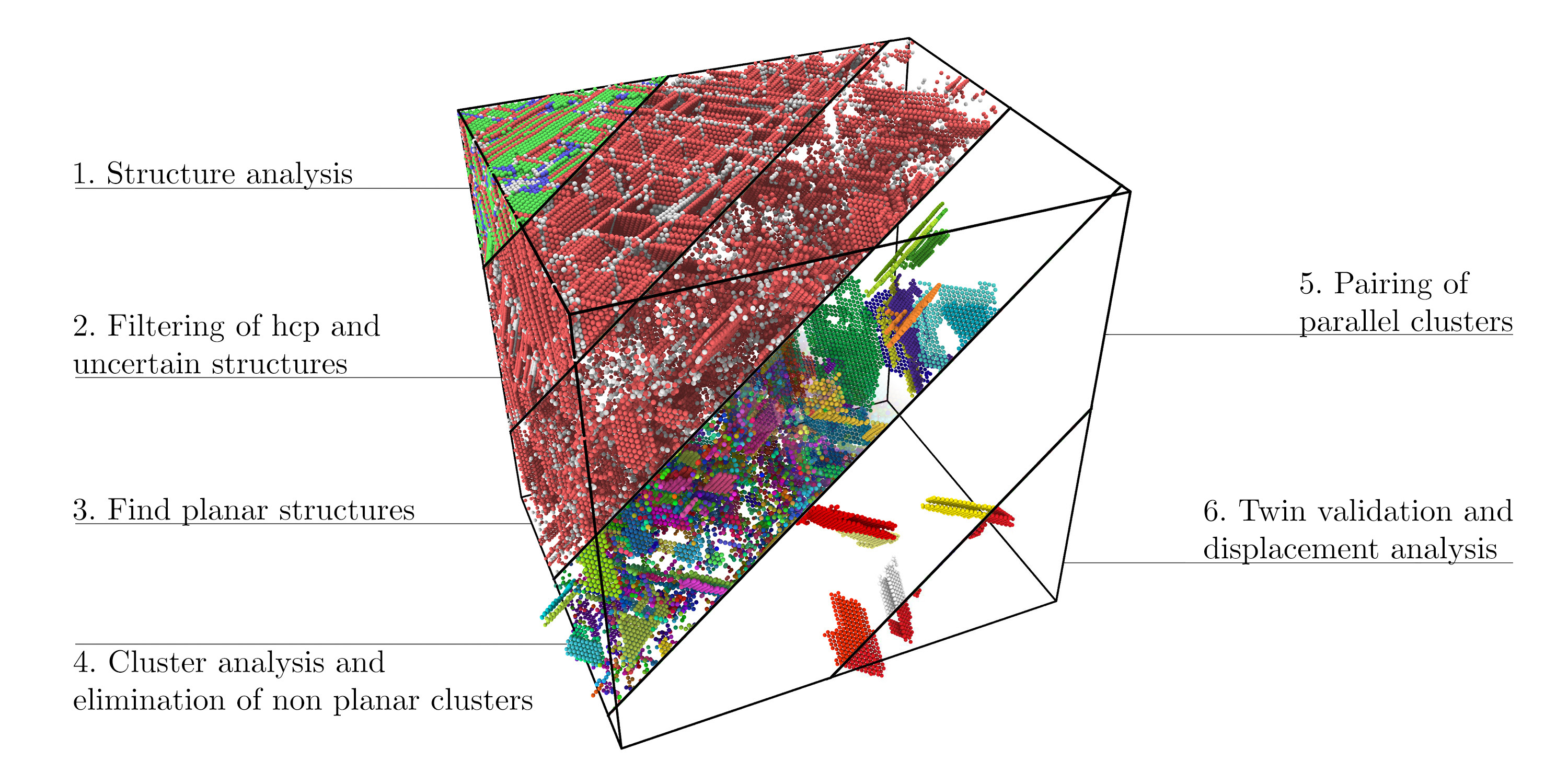}
    \caption{Schematic break-down of the tool processes.}
    \label{fig:GraphicalBreakdown}
\end{figure*}

The utilization of this tool is divided into two parts.
First, it includes an automated script that acts on a range of given files containing atom or geometry data, e.g., LAMMPS dump files (see Fig.~\ref{fig:GraphicalBreakdown}, 1.-5.).
It builds on the problem of distinguishing between single-layer and all other hcp structures, in particular double hcp layers mentioned in section~\ref{sec:MandM}.
These formations are identified, parameterized, and tracked over time.
The obtained data is visualized or stored for further analysis.

The second part is an interactive OVITO session in which detected twin boundary pairs can be examined in detail (see Fig.~\ref{fig:GraphicalBreakdown}, 6.).
It serves as validation for all detected potential twins, since the occurrence of structures, erroneously identified as twins, cannot be entirely excluded.
For twin boundary pairs verified as true twins, the relative shift between the outside fcc regions can be computed, based on the atomic motion of the environment relative to a previous time step.
In situations involving non-uniaxial loading, this may offer insight into the dominant stress direction that initiated its formation.

\subsection{Automated Python script}

The script makes use of some of OVITO's predefined modifiers and is complemented by custom Python functions.
Algorithm~\ref{alg:findTB} shows how clusters of possible twin boundaries are found utilizing basic OVITO modifiers and a short custom selection script.
It receives a series of files and applies modifiers as follows. 

\DontPrintSemicolon
\SetKwComment{Comment}{}{}
\SetKw{Apply}{Apply:}
\SetKwInOut{Input}{input}\SetKwInOut{Output}{output}
\begin{algorithm}
\caption{Find possible twin boundary atoms}\label{alg:findTB}
\Input{OVITO compatible file with atom data} 
\Output{List of clusters with planar structure}
\BlankLine
\Apply{polyhedral template matching}\;
\BlankLine
\If{structure type is hcp or unidentified}{
    \BlankLine
    \Apply{coordination Analysis} \;
    \BlankLine
    \uIf{unidentified atom connects two well-defined planes}{
        \BlankLine
        append to selection\;
        \BlankLine}
    \ElseIf{hcp atom is part of well-defined plane}{
        \BlankLine
        append to selection \;
        \BlankLine}
    
     \If{selected}{
        \BlankLine
        \Apply{cluster analysis}\;
        \BlankLine
        \If{cluster size > threshold}{
        \BlankLine
         add to list of planar structures\;}
         }       
        }

\end{algorithm}

The structure type property is determined using polyhedral template matching (PTM)~\cite{Larsen_2016}, currently being the most reliable (and robust at high temperature~\cite{ma14010060} and strain~\cite{EDER2022101588}) structure identification algorithm, and the only one that has the capability to determine lattice orientation.
Information about the orientation is needed to analyze the lattice environment of a suspected twin to confirm or reject its designation.
Next, the coordination number for every atom of hcp-like or unidentified structure is computed, ignoring all other atom types (they are not considered as neighbors).
The coordination number is determined by counting the neighboring atoms within a given cutoff distance.
This cutoff value, which also applies to the cluster analysis, can be determined by evaluating the atoms' radial distribution function (RDF).
The RDF is a measure of the probability density of finding two particles at a given distance in an atomic system.
The RDF curve shows sharp peaks at certain radial distances, indicating the presence of well-defined interatomic distances for the first, second, and $n$-th neighbors~\cite{RDF_accuracy}.
With increasing temperature, the peaks become blurred, but can still be automatically analyzed to find the cutoff distance, which is the RDF minimum between the peaks of the first and second nearest neighbors.

In fcc lattices, the densely packed planes are of type \{111\}, and these also become twinning planes in twins.
Any atom located within such a layer has six nearest neighbors in that (perfect) plane. 
Large portions of the structures we are looking for are defined by the above criterion. 
However, because of the deformation twins' nature and distortion of the matrix due to strain, the formations often exhibit steps in otherwise planar structures, and they may also become warped.
This partly results in atoms not being identified as twin boundary atoms and, e.g., in the case of steps, two separate planes are identified instead of one continuous twin boundary.
To be able to successfully include the twin boundary in full, atoms of unidentified structure type are used to connect clusters of well-defined layers in close proximity. 
These linking atoms are defined by having at least two nearest neighbors of structure type hcp, and at least one with a coordination number of 6.
If even one nearest neighbor exceeds a coordination number of 6, the center atom is discarded. 
Atoms of well-defined layers are associated with having a coordination number of 6 and structure type hcp as well as having a minimum of 3 nearest neighbors with the same characteristics.

\begin{figure*}[ht]
    \begin{subfigure}[ht]{0.45\textwidth}
        \centering
        \includegraphics[width=\textwidth]{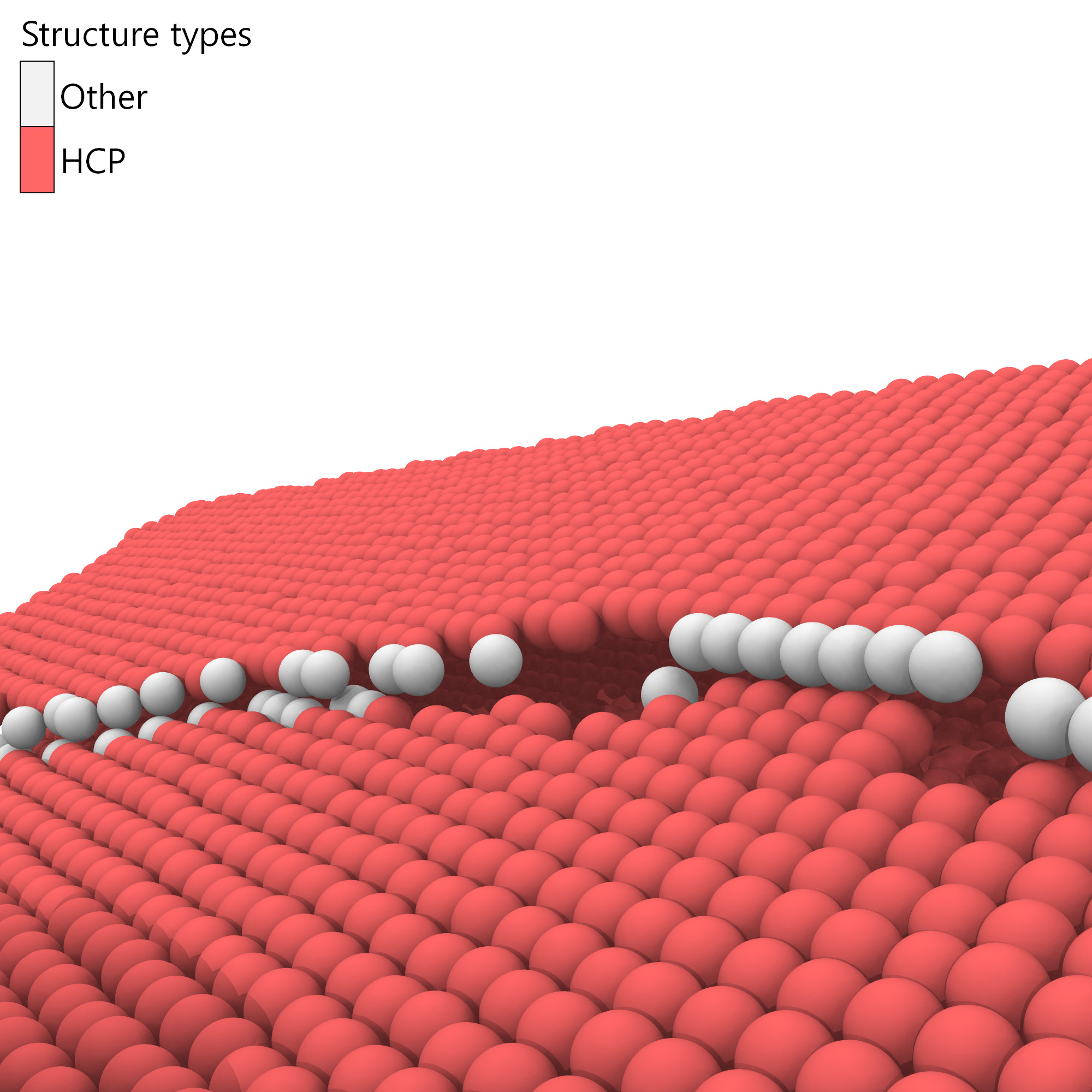}      
        \caption{}
    \end{subfigure}
    \hfill
    \begin{subfigure}[ht]{0.45\textwidth}
        \centering
        \includegraphics[width=\textwidth]{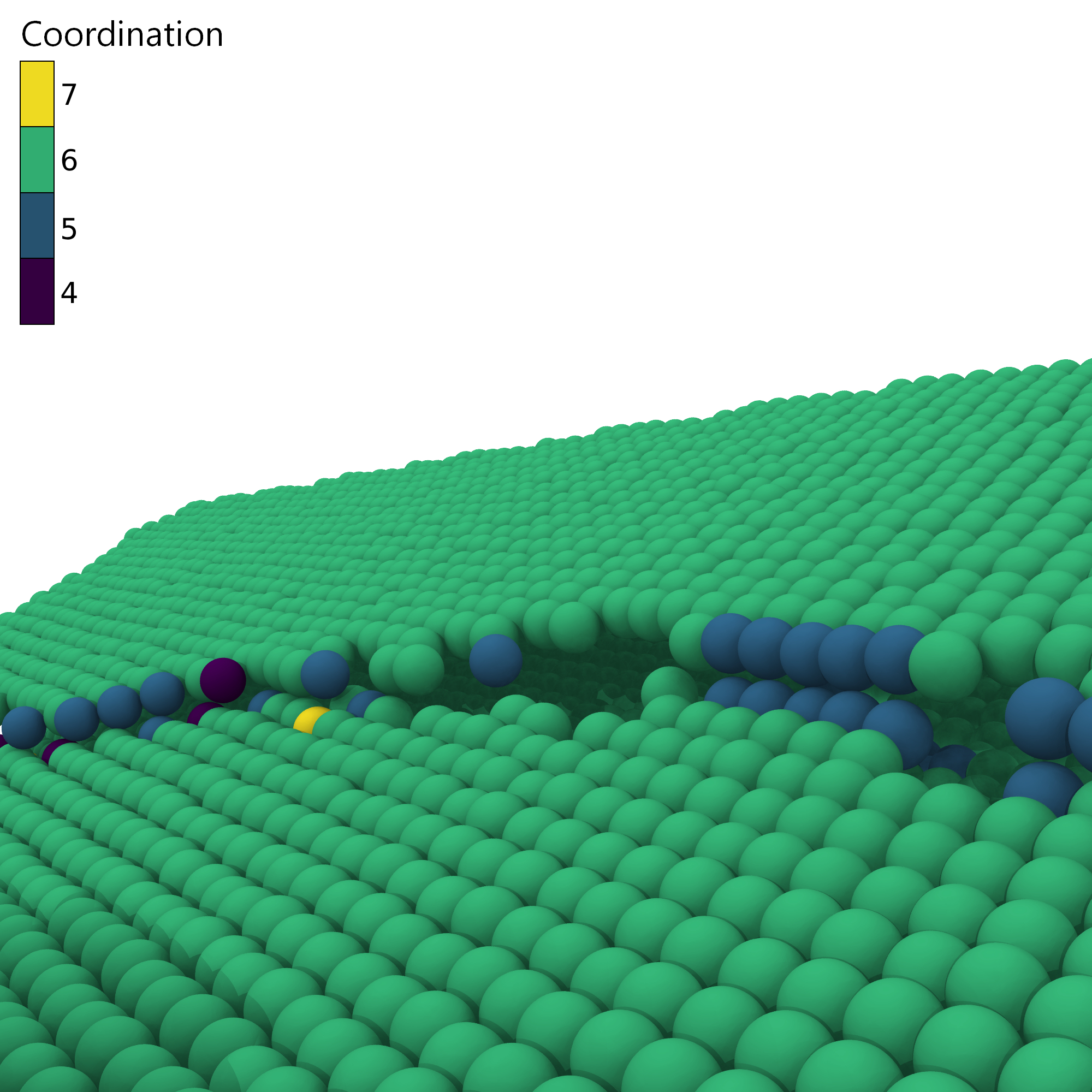}       
        \caption{}
    \end{subfigure}
    \caption{Visualization with focus on the linking atoms between two adjacent planes. a) the planes consist only of hcp type atoms, the linking atoms are primarily of unidentified (``other") structure type. b) in-plane atoms have a coordination of 6; that of the linking atoms varies from 4 to 7. The combination of this information allows to identify the linking atoms, so that the cluster analysis correctly connects both planes.}
    \label{fig:str&coor}
\end{figure*}

Figure~\ref{fig:str&coor} shows a step between two well-defined planes with unidentified atoms and holes (i.e., out-of-plane or non-hcp atoms) in between, where the former act as links between both planes.
In cases where other formations such as double- or multi-layer stacking faults, undefined hcp-like structures, and other crystal defects intersect or border ideal plane configurations, the resulting selection may not be perfectly planar.
This lack of perfect planarity can have a negative impact on the quality of the parametrization process' results.
Specifically, if holes within the selection grow too large, it is possible for coherent twin boundaries to become disconnected.
OVITO's cluster analysis modifier finds disconnected groups (so-called clusters) based on a distance cutoff between single atoms. 
The result is disturbed by the noise from single-atom, non-planar, or one-dimensional (string of atoms) clusters.
This disturbance can be reduced by discarding all clusters with a size of less than a threshold number of atoms (default value: 50).
If one intends to use the temporal tracking feature, the threshold should be kept high, discarding the numerous small twins, as their spatial density heavily affects the tracking accuracy. 
The removal of unwanted structures not covered by this will be performed at a later stage.
Thereby, every single layer and its corresponding atoms are tagged and sorted into a list according to size.
In case of periodic boundary conditions, clusters straddling the periodic boundaries must be unwrapped, meaning their atoms are partly re-positioned.
This is necessary to ensure a correct parameterization of the plane.
However, clusters outside the simulation cell will no longer have a connection to any atoms discarded earlier.
Thus, twin validation and displacement analysis are not available for these objects, since the re-positioned atoms have no adjacent fcc neighbors whose orientation could be analyzed.

\begin{algorithm}[ht]
\caption{Parameterize planes}\label{alg:paramPl}
\Input{List of clusters with supposed planar structure} 
\Output{Data table containing plane-defining vectors}
\BlankLine

\BlankLine
\For{all clusters}{
    principal component analysis\;
    \eIf{shape not planar}{
    delete cluster from planar structures list\;
    
    }{save normal vector and one orthogonal eigenvector to a list}
}
\end{algorithm}

\begin{figure}
    \centering
    \includegraphics[width=65mm]{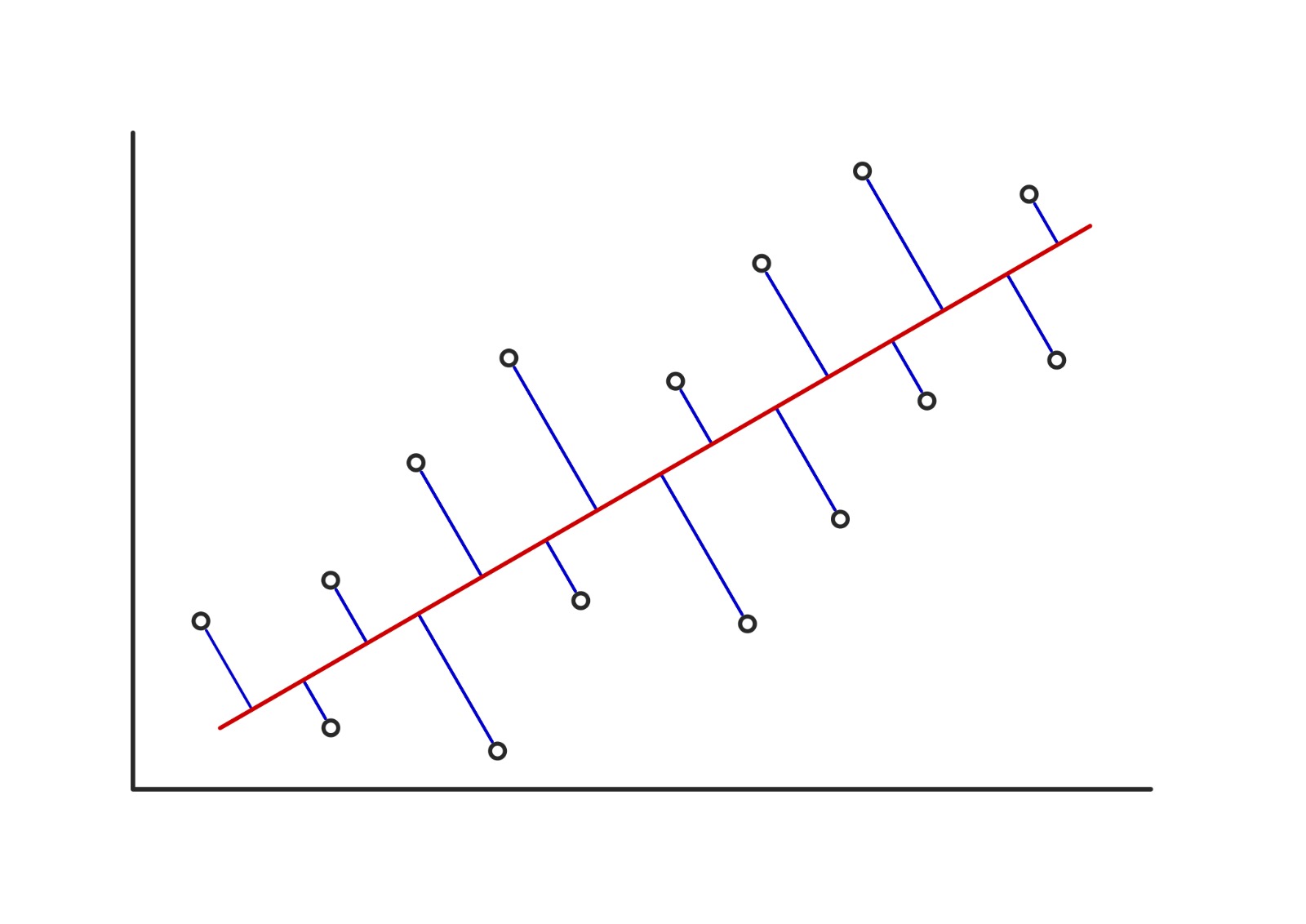}
    \caption{Regression line in two dimensions computed with principal component analysis (adapted from \cite{pcaVSmlr}).}
    \label{fig:PCA2d}
\end{figure}

In the following, regression planes are calculated based on each cluster's atom positions.
Algorithm~\ref{alg:paramPl} summarizes how this helps to eliminate any further shapeless clusters and to allow further processing of mathematically described planes.
Principal component analysis (PCA) is an unsupervised method with all independent but still correlated variables~\cite{Jolliffe2002}, which results in the desired orthogonal regression (see Fig~\ref{fig:PCA2d}). 

PCA essentially causes a principal axis transformation, finding a more reasonable basis of a vector space for a certain dataset.
With the origin at the center of the point cloud, the axes of the new coordinate system are rotated successively in the direction of the next largest variance~\cite{Hauptkomponentenanalyse}.
Generally, the first few principal components will account for most of the variation in the original variables if a set of more than two variables shows substantial correlations among them~\cite{Jolliffe2002}. 
Transferred to the spatial coordinates at hand, this allows a rough estimate of the shape of the point cloud. 
Since the studied twin boundary atoms exhibit a strong linear correlation regarding their positions in the direction of the plane-spanning vectors, PCA can be successfully applied to parameterize planes.

In practice, the application part is limited to setting up the covariance matrix~\cite{nla.cat-vn189234} and computing its eigenvalues and eigenvectors. 
These resemble the variances in atom positions and their directions, respectively.
From the above stochastic and mathematical relationships, the approximate geometry of a parameterized cluster can be obtained.
Since a perfect plane has the minimal extent, thus variance, in the direction of its normal vector, this plane-defining vector is the eigenvector corresponding to the smallest eigenvalue. 
The remaining two eigenvalue-eigenvector-pairs allow the plane's aspect ratio to be determined, which facilitates the elimination of non-planar structures.

\begin{figure}[ht]
    \centering
    \includegraphics[width=90mm]{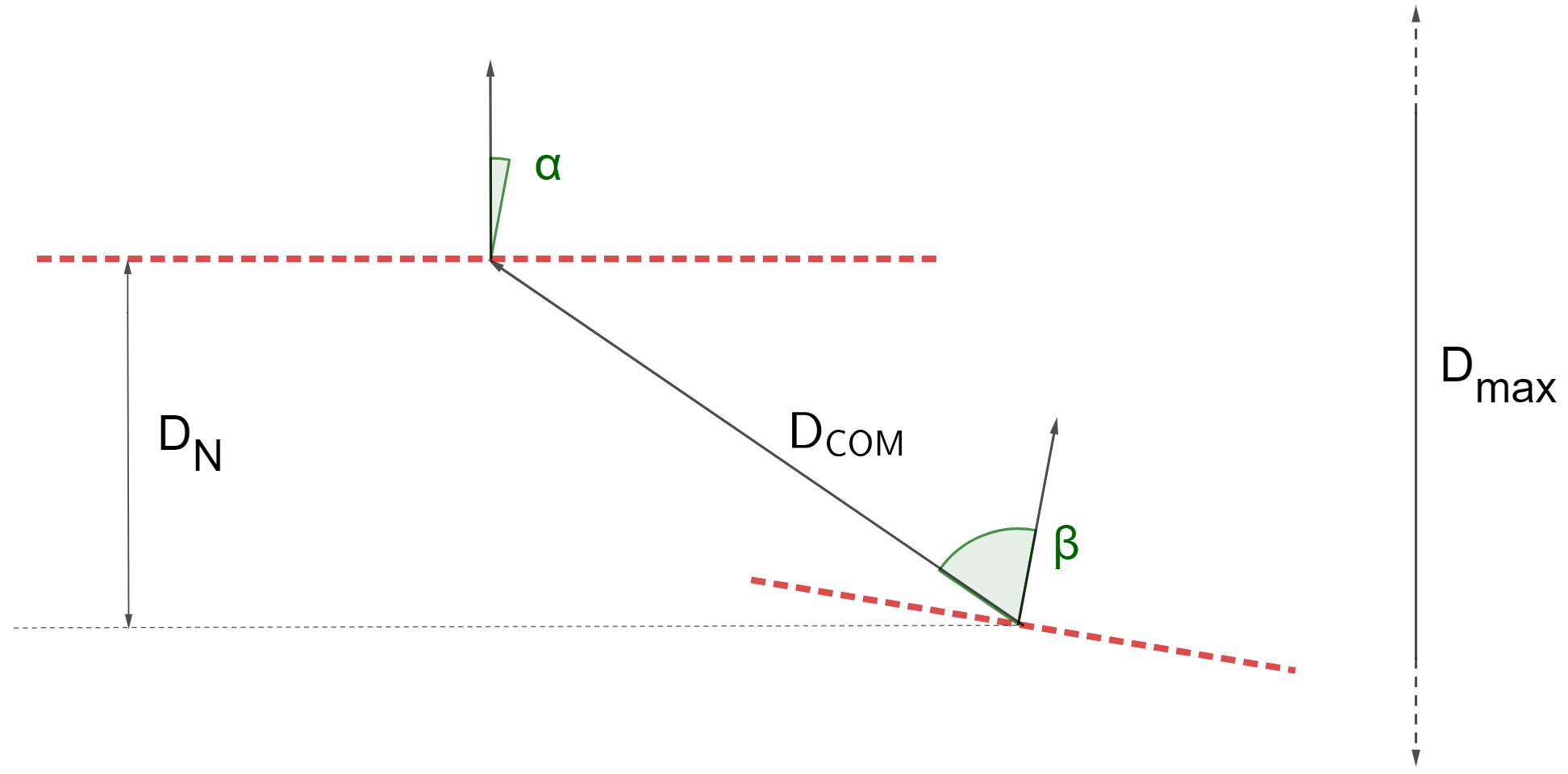}
    \caption{2-dimensional visualization of geometrical relations computed for pair finding.}
    \label{fig:PairingParam}
\end{figure}

The obtained list of possible twin boundary forming clusters is now unfolded into a 2-D array, comparing the clusters individually.
The following geometrical relations (for abbreviations, see table~\ref{tab:planeParams}) are calculated for each comparison (see Fig.~\ref{fig:PairingParam} for a visualization):
\begin{table}[ht]
    \centering
    \caption{Abbreviations for geometrical relations computed between two plane-shaped clusters.}
    \begin{tabular}{c|p{9cm}}
        $\alpha$ & angle between normal vectors \\
        $\beta$ & angle between one normal vector and each cluster's \\ 
        & center of mass (COM) connecting vector\\
        $D_N$  & distance in normal vector direction between COMs\\
        $D_{COM}$ & distance between COMs\\
        $S_{rel}$ & size relation (ratio between number \\
        & of atoms per cluster)\\
        $D_{max}$ & largest dimension of the simulation cell
    \end{tabular}
   
    \label{tab:planeParams}
\end{table}

\begin{table}[ht]
    \centering
    \caption{Empirically acquired conditions for cluster pairing algorithm~\ref{alg:findplanes}}
    \begin{tabular}{c|p{7cm}}
        $C_{\alpha}$ & $\alpha < 5\degree$  \\
        $C_{\beta}$ & $\beta < 90\degree-90\degree\times20 D_N/D_{max}$\\
        $C_{COM}$ & $D_{COM}<3 \times D_N$\\
        $C_{Size}$ & $0.2<S_{rel}<5$
    \end{tabular}
    \label{tab:PairingConditions}
\end{table}

Table~\ref{tab:PairingConditions} displays the empirically obtained conditions used in the plane pairing algorithm~\ref{alg:findplanes}.
The tolerance in condition $C_\alpha$ is necessary because of unavoidable deviations from a perfect plane configuration (single atoms) and the consideration of stress-related twin boundary curvature.
The remaining conditions reduce the number of crystallographically unreasonably paired planes.
$C_\beta$ prevents that two distant planes, which in reality could never form a twin, are paired (if $D_N$ is not small against $D_{max}$).
$C_{COM}$ forbids that two different clusters in the same plane are assigned to each other.
Lastly, $C_{Size}$ provides a balanced size relation between twin boundaries. 

\begin{algorithm}[ht]
\SetKwComment{Comment}{}{}
\SetKw{Compute}{Compute:}
\SetKw{Assign}{Assign:}
\SetKw{export}{Export:}
\caption{Find parallel planes}\label{alg:findplanes}
\Input{List of normal vectors of clusters with planar structure} 
\Output{LAMMPS dump file and data tables}
\BlankLine
\BlankLine
\For{all permutations of the computed planes}{
\BlankLine
    \Compute{$\alpha$, $\beta$, $D_N$, $D_{COM}$, $S_{rel}$}  \hspace{35mm}\Comment{//see Table~\ref{tab:planeParams}}
    \BlankLine
    \If{plane combination satisfies all conditions  \hspace{21mm}\Comment{//see Table~\ref{tab:PairingConditions}}}{           
    \BlankLine
    \eIf{plane already paired}{
    \BlankLine
    \If{$D_{COM}$ (selected pair) < than $D_{COM}$ (already paired)}{
    \BlankLine
    replace previously found pair with selected pair 
    }
    }{append plane pair to list of possible twin boundary pairs}
    }
}
\Assign{``possible twin" marker to atoms of selected pairs}\;
\export{related data tables} 
\end{algorithm}

All important values of the found pairs are exported in case the user decides to do an interactive OVITO session afterwards.
If multiple matches are identified, a comparison of the values of $D_N$ is performed for each of the matches. The two matches with the closest values of $D_N$ are then selected as the final result.\\

The tracking algorithm~\ref{alg:trackTwins} relies on comparison of position and orientation between timesteps.
The mean values of the COM and normal vector properties of the two twin boundaries are compared between the currently detected structures and the information stored from the previous timestep.
At the first provided frame, the list of detected twin structures is initiated and will serve as reference for the next frame.
Starting at frame 2, the data table containing information from the previous timestep is read.
A twin is tracked if a structure from the referenced timestep is of the same orientation (5° tolerance), less than 15~\AA\ (based on a sampling interval of 2~ps), away in normal direction, and if the combined COM lies within a range of 50~\AA.
The previous twin's ID will be assigned to the successfully tracked twin.
If no match is found, the selected twin receives its own unique ID.
After iterating through all structures of the current timestep, the timestep of origin and of disappearance of non-tracked twins are transferred to the updated data table.
In case of multiple matches, the one closest to the reference position is selected.
This tends to be the most computationally demanding aspect, depending on the number of found clusters.

\SetKw{export}{Export:}
\SetKw{Tracking}{Tracking}
\SetKw{NewTwin}{New twin}
\begin{algorithm}[ht]
    \caption{Track twins over provided timesteps}\label{alg:trackTwins}
    \Input{List of positional data of twins detected in previous timestep} 
    \Output{Altered atom data file and data table of tracked, lost, and new twins}
    \BlankLine
    \For{all twins found at current timestep}{
    \uIf{orientation and position similar to twin in previous timestep}{
        \BlankLine
        \Tracking\;
        \BlankLine}
    \ElseIf{no match found}{
        \BlankLine
        \NewTwin\;
        \BlankLine}
        }
    carry over data of lost twins\;
    \export{LAMMPS dump file and updated tracking data table}
\end{algorithm}


\subsection{Interactive OVITO session}
\label{sec:interactive}
\begin{figure*}[ht]
     \centering
     \begin{subfigure}[t]{0.65\textwidth}
         \centering
         \includegraphics[width=\textwidth]{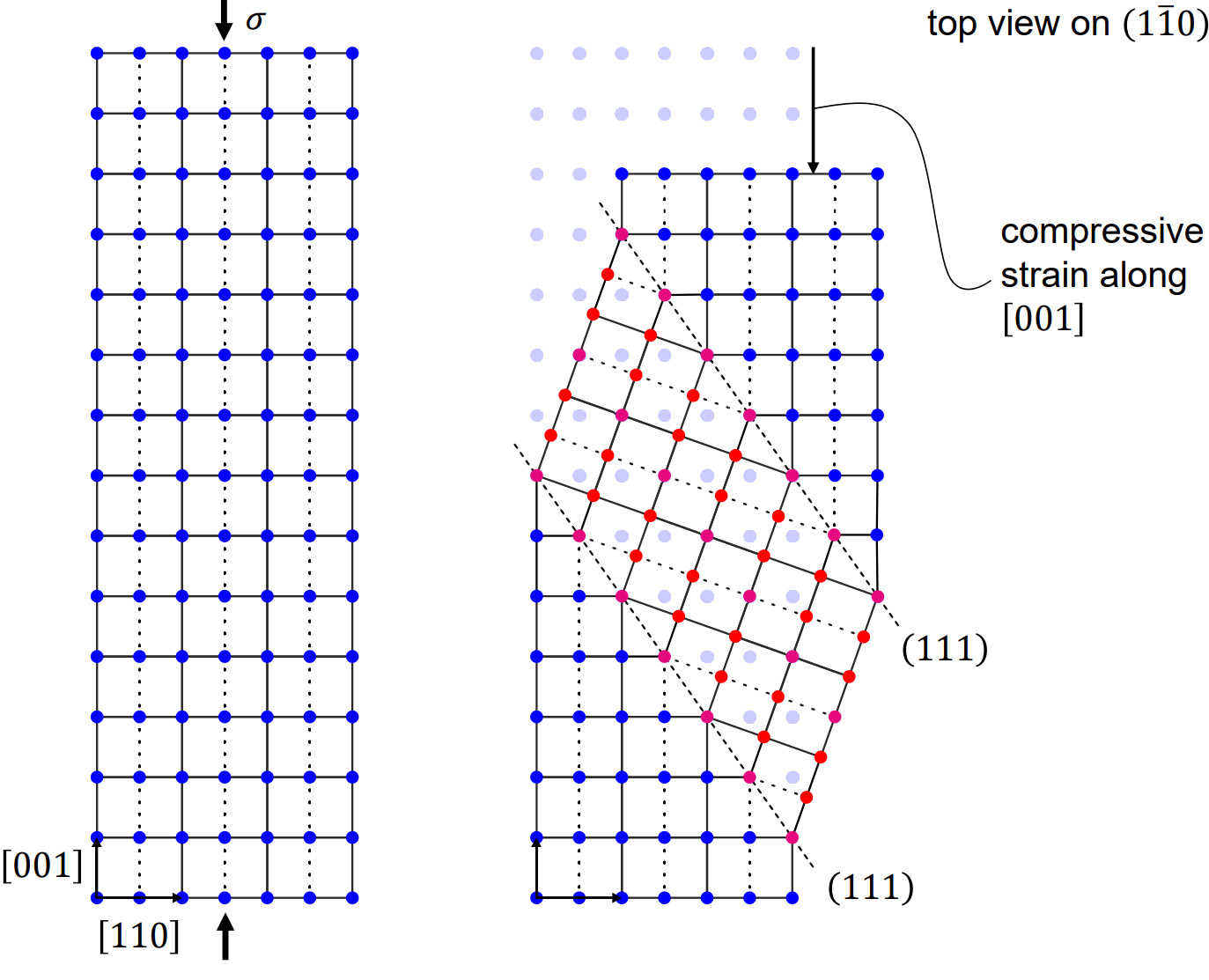}
         \caption{}
         \label{fig:GitterDruck}
     \end{subfigure}
     \hfill
     \begin{subfigure}[t]{0.34\textwidth}
         \centering
         \includegraphics[width=\textwidth]{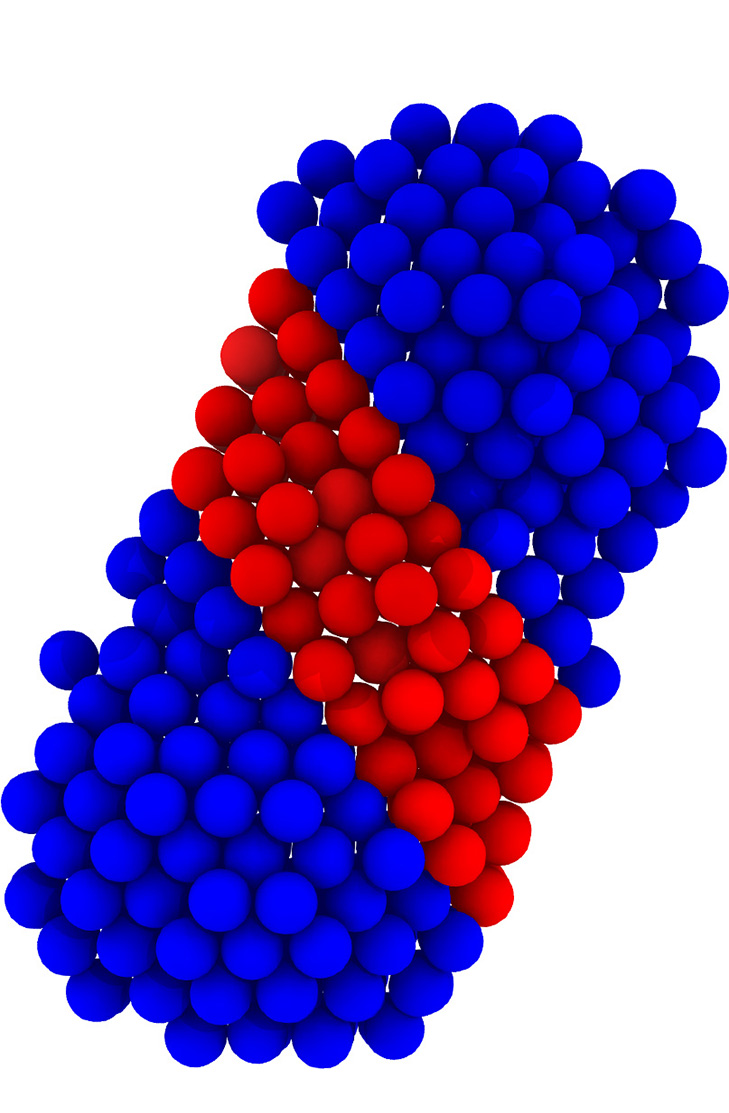}
         \caption{}
         \label{fig:3dGitter}
     \end{subfigure}
    \caption{Visualized concept of dividing the selection into distinct regions, which facilitates the orientation and displacement analysis: 
    relative displacements of the blue regions after twinning as an indication of the type of formation. 
    (a) 2-dimensional scheme of compression twin formation in fcc lattices~\cite[p. 4]{KauffmannPresentation}. 
    (b) Atom selection according to algorithm~\ref{alg:speeradaptive}: 
    3-dimensional equivalent to figure (a): twin atoms (red) and matrix atoms (blue).}
    \label{fig:ZugDruckBestimmung}
\end{figure*}

While the successfully paired planes have passed the first requirement for twin identification, they have not yet undergone validation as genuine twins.
The results may still contain erroneously paired twin boundaries or structures that do not qualify as twin boundaries at all.
Therefore, further identifiers such as the associated changes in lattice orientation and other related characteristics must be confirmed.
To address this, each twin boundary pair is systematically re-evaluated.
For identification of the crystallographic orientations in the twins' immediate vicinities, the results of the first script can be passed into the OVITO data pipeline.
OVITO can load either a single file or an entire sequence. 
In both cases, all required data tables are automatically imported for each timestep. 
The evaluation of previously found potential twin boundary pairs is first done by examining the orientation of the areas between and to some extent also outside the recognized planes. 
The difference in crystallographic orientation between the twin and the matrix atoms can be seen in figure~\ref{fig:GitterDruck} (in this case the twin is formed during compression).
This well known schematic representation of a deformation twin, served as inspiration to apply analogous segmentation to the detected twin boundary pairs (see Fig.~\ref{fig:3dGitter}).
For this purpose, the orientation of all fcc-type and twin boundary atoms inside a cylinder aligned with the vector connecting the centers of mass are saved together with their normal distance to one of the planes.
The cylinder's radius is determined by the size of the cluster, but has a minimum of 10~\AA\, to compensate for possible holes and small defects between planes.


Because of the twin forming in parallel planes, clusters of similar crystallographic orientation appear at regular intervals when the collected data is plotted.
Every cluster represents one parallel layer that the cylindrical mask has intersected.
Only if steps of a twin boundary are intersected, the cluster will scatter along the vertical axis, as this layer shares atoms of matrix or inner twin and twin boundary.
The PTM stores the orientation of any analyzed atom as a quaternion, a four-component vector whose entries can be a suitable choice for a scalar identifier of orientation.
 
\begin{figure}[ht]
    \centering
    \begin{subfigure}[t]{0.49\textwidth}
         \centering
         \includegraphics[width=\textwidth]{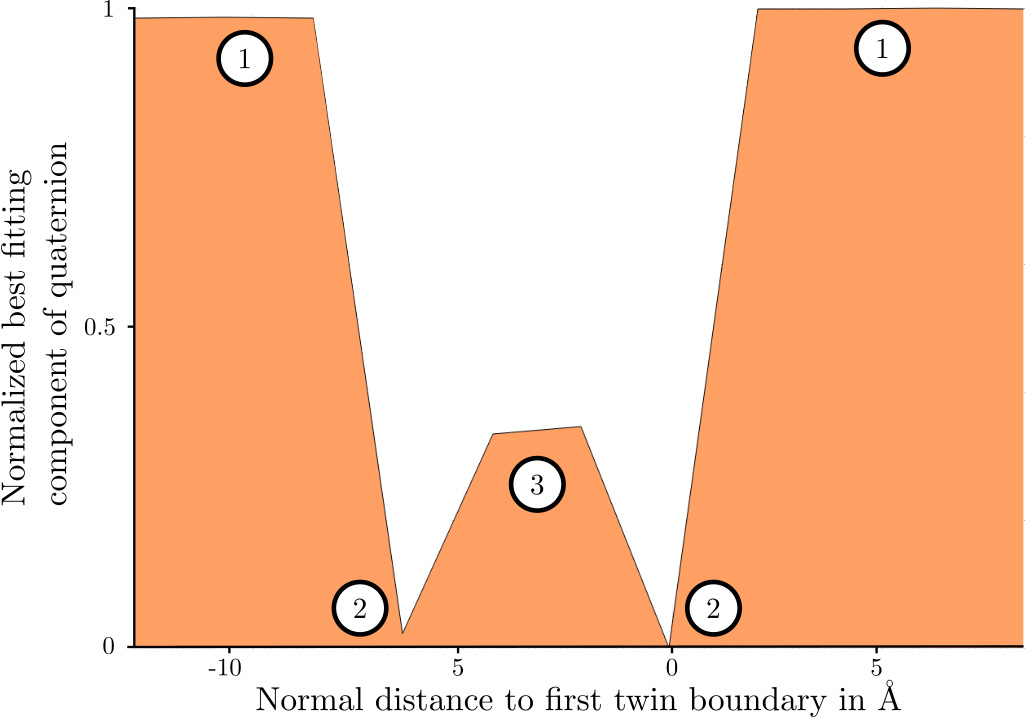}
         \caption{}
         \label{fig:Plot_W}
     \end{subfigure}
     \hfill
     \begin{subfigure}[t]{0.49\textwidth}
         \centering
         \includegraphics[width=\textwidth]{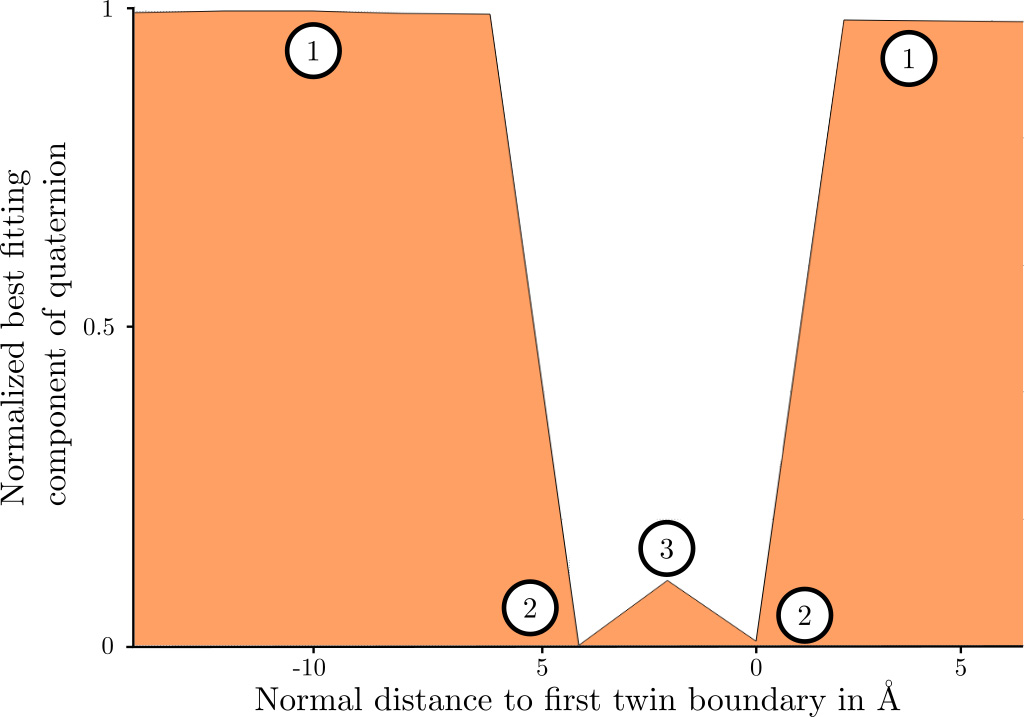}
         \caption{}
         \label{fig:Plot_U}
     \end{subfigure}
  
    \caption{Average crystallographic orientation against normal distance of atom layers parallel to twin boundaries: a) twin fcc structure with distinct difference in orientation to twin boundaries. b) twin fcc structure with little to no difference in orientation to twin boundaries. 1: surrounding matrix, 2: twin boundaries of hcp-like structure, 3: twin fcc structure enclosed by twin boundaries}
    \label{fig:PlotsNumbered}
\end{figure}

A normalization of the resulting scattered orientation data to a range between zero and one helps spread the (possibly small) deviations in orientation between atoms in close proximity.
This enables a more meaningful, high-contrast display of plots, enabling automatic identification of special features (in this context sudden changes in orientation), but it also improves floating point precision.
The aim is to achieve clear demarcation between spatial regions of different orientation.
To determine which component of the orientation quaternion provides the highest contrast between matrix and twin, all four components are compared to the pattern that would be expected for a perfect twin.
For better comparability, these orientation components are then separately "computed as/filtered for their" median values over all atoms in the previously mentioned clusters.
Grouping the data, this time by orientation, makes it possible to autonomously identify the graphs' shape.
Large jumps in the graph come from a significant change in orientation, which ideally indicates the position of the first twin layer, the twin boundary.
In OVITO, the respective atoms are identified as hcp-like and are not connected to the surrounding fcc matrix, which is why in some cases the twin boundary is followed again by a jump in the graph (see Fig.~\ref{fig:PlotsNumbered}).

The two patterns shown in this figure can be divided into five sections of similar orientation in the case of Fig.~\ref{fig:Plot_W} and three for Fig.~\ref{fig:Plot_U}.
All detected structures sharing one of the patterns are reliably identified as true twins. 
In the last step, the pairing algorithm's results are reviewed.
Due to the division of the areas, they can also be considered individually so that knowledge about which atoms are located outside the twin boundaries is gained.

This segmentation is performed according to the quaternion component that has the highest contrast in orientation between matrix and twin regions.
Then, from the median values of the quaternions representing the orientations of these newly defined regions, we can calculate both the rotation angle $\theta$ and the associated axis $\textbf{n}$ between the twinned fcc volume and the surrounding fcc (matrix) region.
If the rotation axis comes out normal to the twinning plane and the rotation angle is close to 60° ($\Sigma$3 boundary)~\cite{OTTE1954349}, the twin has been successfully confirmed.

Both quantities $\theta$ and $\textbf{n}$ can be determined, according to Ref.~\cite{Rollett_Ozturk_2016}, by first forming the (non-commutative) quaternion product of one quaternion with the inverse of a second one, yielding a single quaternion that represents the rotation from one crystal region/orientation into the other:

%
\begin{gather}\label{eq1}
\begin{split}
x \; y	
= & ( x_0 y_0 - x_1 y_1 - x_2 y_2 - x_3 y_3)\\
+ & (x_{0}y_{1}+x_{1}y_{0}+x_{2}y_{3}-x_{3}y_{2})\mathrm {\textbf{i}}\\		
+ & (x_{0}y_{2}-x_{1}y_{3}+x_{2}y_{0}+x_{3}y_{1})\mathrm {\textbf{j}}\\	
+ & (x_{0}y_{3}+x_{1}y_{2}-x_{2}y_{1}+x_{3}y_{0})\mathrm {\textbf{k}} 
\end{split}
\end{gather}
This quaternion is of the form $q=q_0+\textbf{i}q_1+\textbf{j}q_2+\textbf{k}q_3$. 
The corresponding rotation matrix can then be obtained as
\begin{equation}\label{eq:rotmat}
\textbf{Q}=
\begin{bmatrix}
1-2q_2^2-2q_3^2 & 2q_1q_2+2q_0q_3 & 2q_1q_3-2q_0q_2\\
2q_1q_2-2q_0q_3 & 1-2q_1^2-2q_3^2 & 2q_2q_3+2q_0q_1\\
2q_1q_3+2q_0q_2 & 2q_2q_3-2q_0q_1 & 1-2q_1^2-2q_2^2
\end{bmatrix} .
\end{equation}
From this, the rotation axis $\textbf{n}$ is calculated component-wise using the non-diagonal elements $a_{ij}$ of $\textbf{Q}$.
\begin{equation}
    \textbf{n}=\frac{(a_{23}-a_{32}),(a_{31}-a_{13}),(a_{12}-a_{21})}{\sqrt{(a_{23}-a_{32})^2+(a_{31}-a_{13})^2+(a_{12}-a_{21})^2}}
\end{equation}
The diagonal elements of $\textbf{Q}$ are used to compute the rotation angle $\theta$.
\begin{equation}
    \theta = \arccos{0.5(\mathrm{tr}(\textbf{Q})-1)}
\end{equation}

\begin{algorithm}[ht]
    \caption{Validation and  displacement analysis}\label{alg:speeradaptive}
    \Input{Atom data files of previous algorithm} 
    \Output{Orientation-vs.-normal-distance plots}
    \BlankLine
    \For{every twin boundary pair}{
        \BlankLine
        select atoms within cylindrical mask\;
        \BlankLine
        find best fitting quaternion component\;
        \BlankLine
        plot orientation against normal distance to one plane\;
        \BlankLine
        \If{graph matches pattern}{
        \BlankLine
            compute rotation angle and axis for validation\;
            \BlankLine
            get displacement of regions outside the twin\;
            \BlankLine
            \eIf{distance between region's COMs decreased}{
            \BlankLine
            mark as narrowing twin}
            {\BlankLine
            mark as widening twin}
        }
    }
\end{algorithm}

\begin{figure}[ht]
    \centering
    \begin{subfigure}[t]{0.49\textwidth}
        \includegraphics[width=\textwidth]{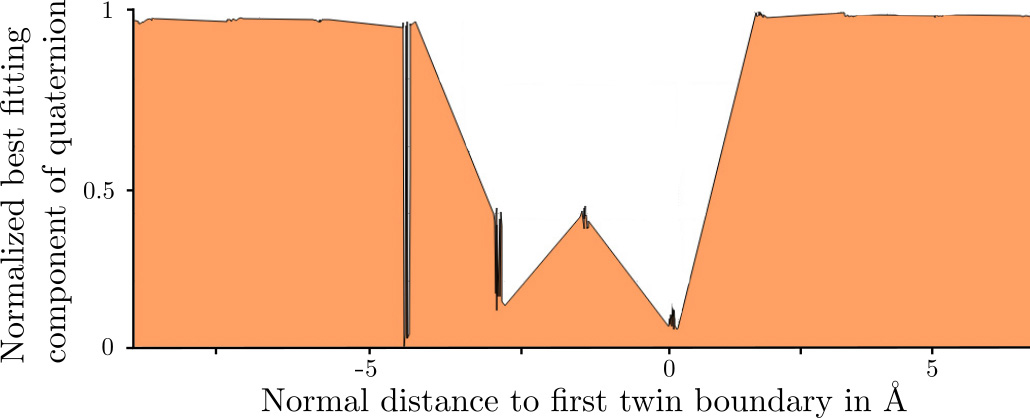}
         \caption{Cutoff radius of 5~\AA.}
         \label{fig:Cutoff5}
    \end{subfigure}
     \hfill
     \begin{subfigure}[t]{0.49\textwidth}
         \centering
         \includegraphics[width=\textwidth]{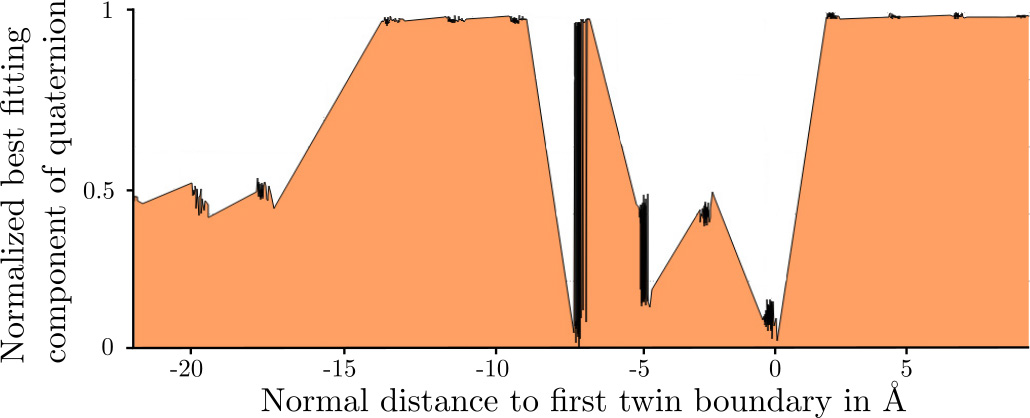}
         \caption{Cutoff radius of 10~\AA.}
         \label{fig:Cutoff10}
     \end{subfigure}
      
     \begin{subfigure}{0.49\textwidth}
         \centering
         \includegraphics[width=\textwidth]{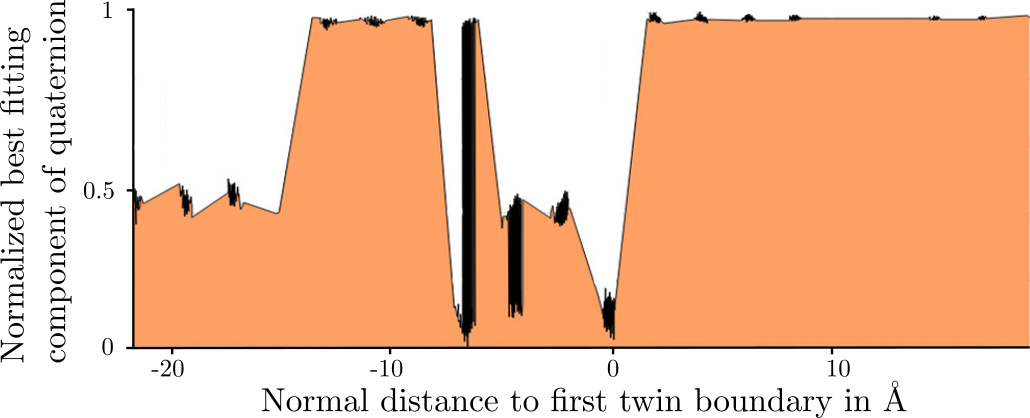}
         \caption{Cutoff radius of 15~\AA.}
         \label{fig:Cutoff15}
     \end{subfigure}
     \hfill
     \begin{subfigure}{0.49\textwidth}
         \centering
         \includegraphics[width=\textwidth]{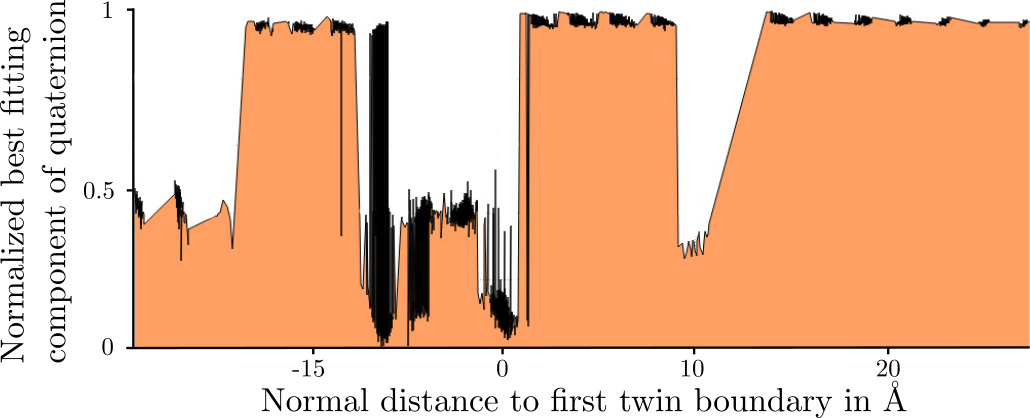}
         \caption{Cutoff radius of 25~\AA.}
         \label{fig:Cutoff25}
     \end{subfigure}
    \caption{Various cutoff radii for the selection mask during orientation analysis.}
    \label{fig:cutoffradii}
\end{figure}

It is obvious that the success of the orientation analysis depends significantly on the selection of atoms.  
Modifying the cutoff radius of the sphero-cylindrical mask used for the selection brings only visual improvements.
Figure~\ref{fig:Cutoff5} suggests an advantage of smaller cutoffs, as the drop in orientation on the left is avoided (compared to the other plots).
Since this is highly individual though, and the tolerance for outliers is lowered, a minimum cutoff of 10~\AA\, was selected.
Increasing the value to 15~\AA\, shows no remarkable improvement (compare Figs.~\ref{fig:Cutoff10} and~\ref{fig:Cutoff15}).
However, it is certain that an overly large radius will exacerbate the discontinuity (see in particular the region between the twin boundaries in Fig.~\ref{fig:Cutoff25} and previous plots).
In highly dynamic systems, increasing the cutoff will always deface the plots, as the density of hcp-like structures is high, which mostly account for the arbitrary changes in orientation.

The issue of intersecting steps within twin boundaries during orientation analysis can potentially be addressed by combining information from the entire preceding analysis sequence.
This approach can help consolidate twin validation and determine its relative shift in each system snapshot.
Since the steps resemble the moving Shockley partial dislocations, there will typically be at at least one frame where the plot is unambiguous.

Comparing the displacement of the sections surrounding the twin (see Fig.~\ref{fig:PlotsNumbered} sections 1 and Fig.~\ref{fig:ZugDruckBestimmung} blue atoms), their relative movement can be computed. 
This relative shift, either narrowing or widening, is assigned to every successfully detected and validated twin. 
In non-uniaxial load cases, this might provide the possibility to deduce the predominant sign of stress that led to that twin's formation.
This approach is based on the observable relative movement of the areas outside the twin (see Fig.~\ref{fig:GitterDruck}).
This concept is transferred to the masked region (see Fig.~\ref{fig:3dGitter}) by computing the average position (equivalent to COM) and displacement of both atom groups (see blue atoms) and comparing the resulting average positions of both frames.
If, after subtracting the displacement from the COM for both regions, the distance between the new positions has decreased in comparison to the distance only between the centers of mass, it suggests the outside fcc regions have moved closer together. 
Analogous thoughts apply in the opposite case.
Algorithm~\ref{alg:speeradaptive} summarizes the key points of the interactive session.

\section{Representative Case Study}
\label{sec:Results}

In this section, we present a short case study to showcase the capabilities and benefits of the presented tool.
The object of the study was a copper single crystal with initial dimensions of 25$\times$25$\times$25~nm$^3$ under shear, and an embedded atom method (EAM) potential was used to model the atomic interactions~\cite{Cu_eam_potential}.
\begin{figure}[ht]
     \centering
     \begin{subfigure}[ht]{0.49\textwidth}
         \centering
         \includegraphics[width=\textwidth]{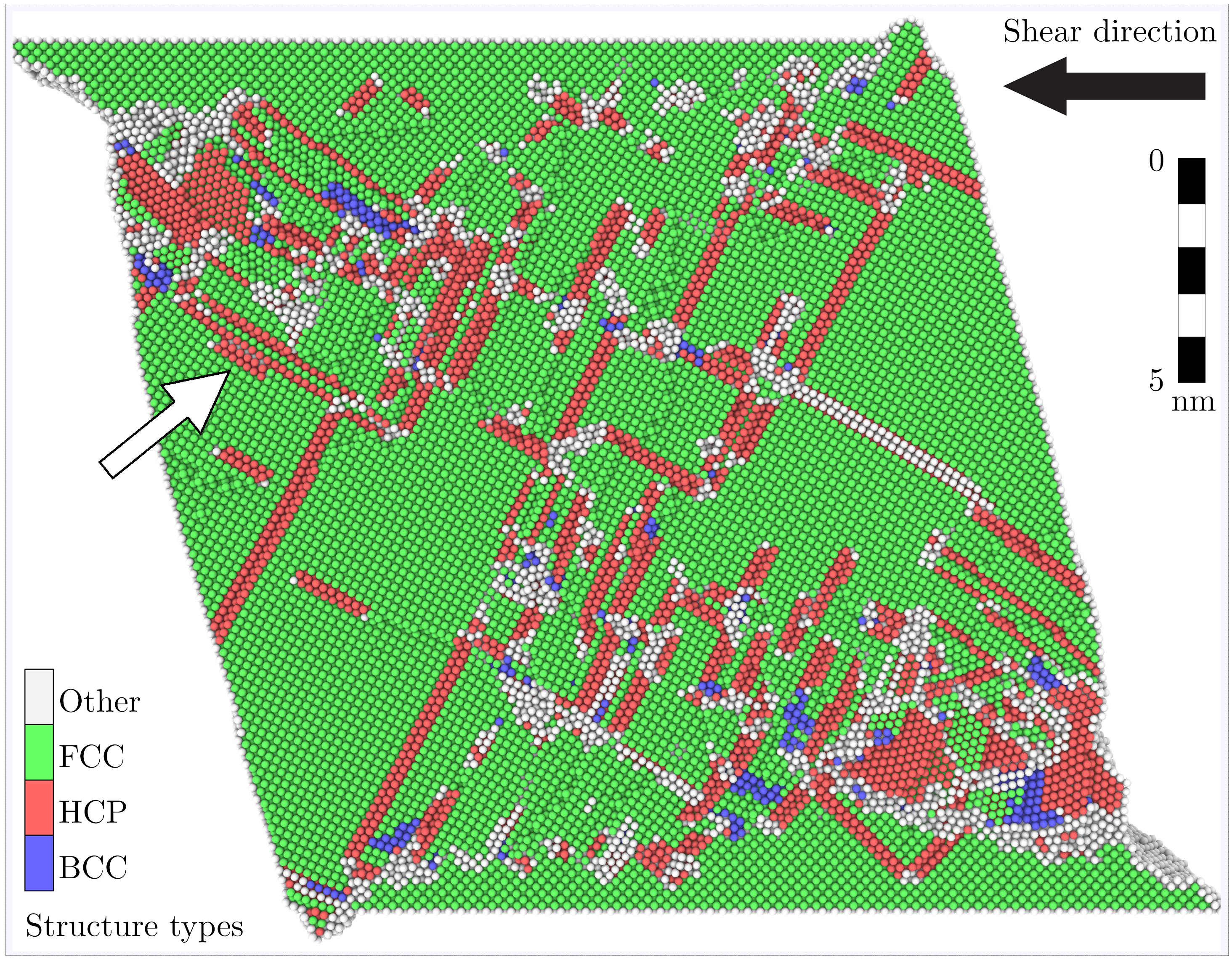}
         \caption{}
         \label{fig:CS_PTM_Schnitt}
     \end{subfigure}
     \hfill
     \begin{subfigure}[ht]{0.49\textwidth}
         \centering
         \includegraphics[width=\textwidth]{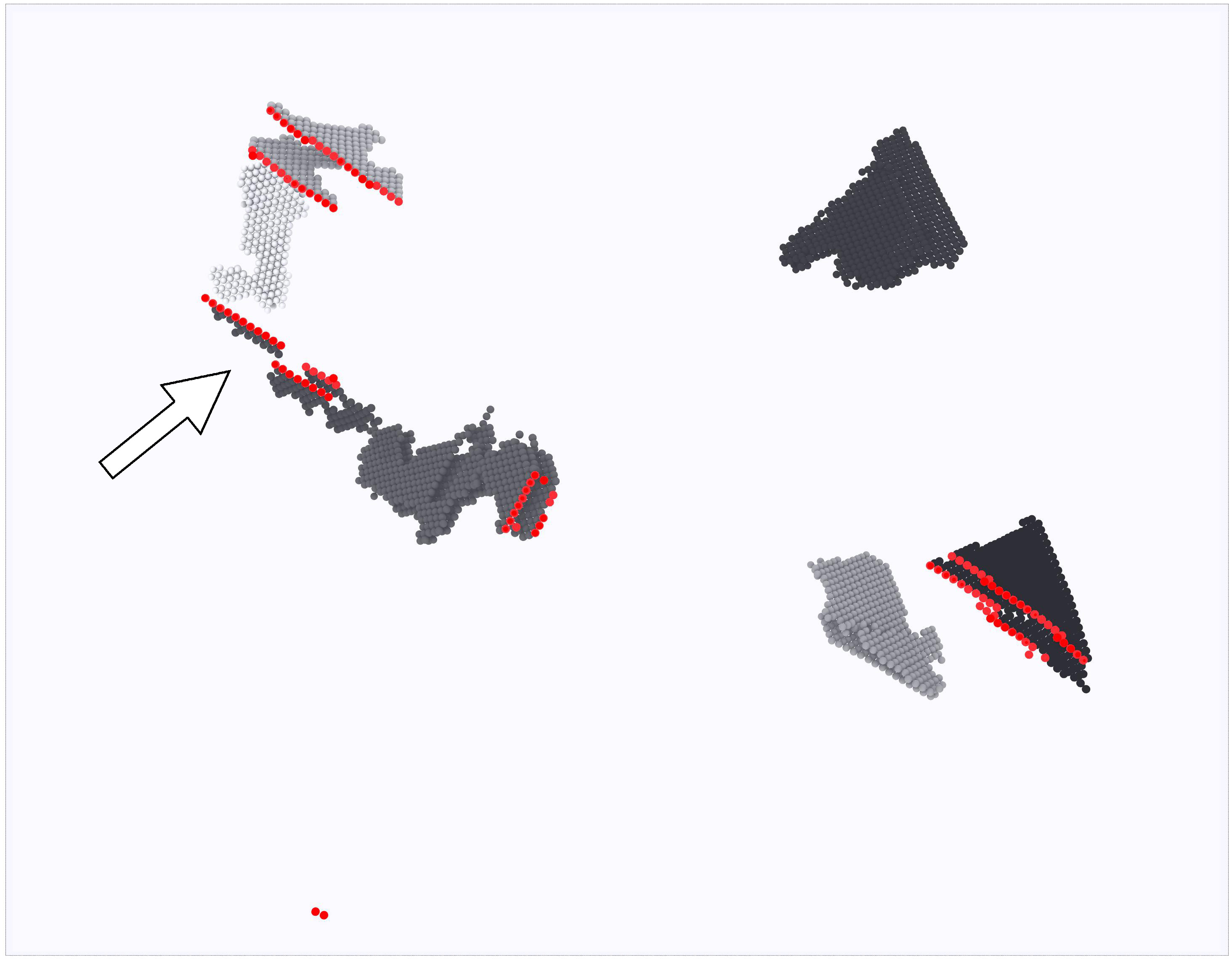}
         \caption{}
         \label{fig:CS_Twin_Schnitt}
     \end{subfigure}
    \caption{Same cross-section of the simulation showing a) the structure identified with PTM and b) the identified twin-like structures with atoms in the slicing plane highlighted in red.}
    \label{fig:CS_schnitte}
\end{figure}
The system was equilibrated at a temperature of 200~K with the $x$, $y$, and $z$ axes corresponding to the [100], [010] and [001] directions, respectively.
A rigid top section consisting of two monolayers of Cu is sheared in positive $x$ direction at a velocity of 100~m/s (corresponding to a strain rate of $\simeq$0.4~ps$^{-1}$).
A velocity gradient was applied to a region of atoms neighboring this mobile layer to introduce the shear into the system more gently. 
The timestep size was set to 2~fs, and the first twin structures were detected after 12~ps.
Figure~\ref{fig:CS_schnitte} gives an overview of which structures are identified as possible twins. 
The center left structure found in both Fig.~\ref{fig:CS_PTM_Schnitt} and~\ref{fig:CS_Twin_Schnitt} (indicated by the white arrow) is noteworthy as it is bordered by a stacking fault, which disfigures the resulting structure.
The presence of atom accumulations within twin boundaries, particularly when situated centrally, is observed to have a limited influence on the direct detection of coherent twin boundaries.
However, these accumulations can significantly perturb the correct pairing (owing to contorted plane parameterization) and hinder the twin validation through atom orientation analysis.

\begin{figure}[ht]
     \centering
     \begin{subfigure}[t]{0.49\textwidth}
         \centering
         \includegraphics[width=\textwidth]{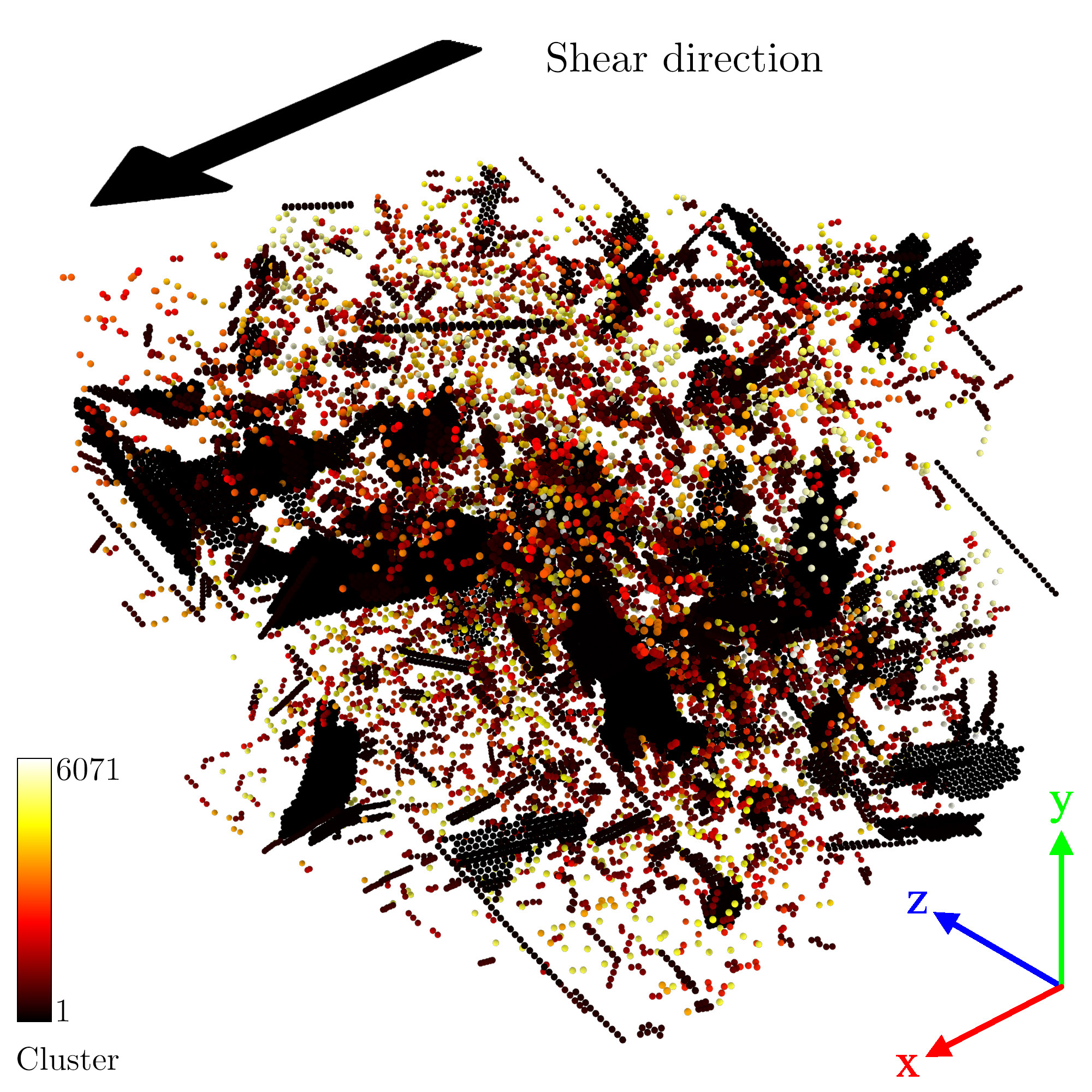}
         \caption{Cluster analysis of detected planar structures}
         \label{fig:CSCLus}
     \end{subfigure}
     \hfill
     \begin{subfigure}[t]{0.49\textwidth}
         \centering
         \includegraphics[width=\textwidth]{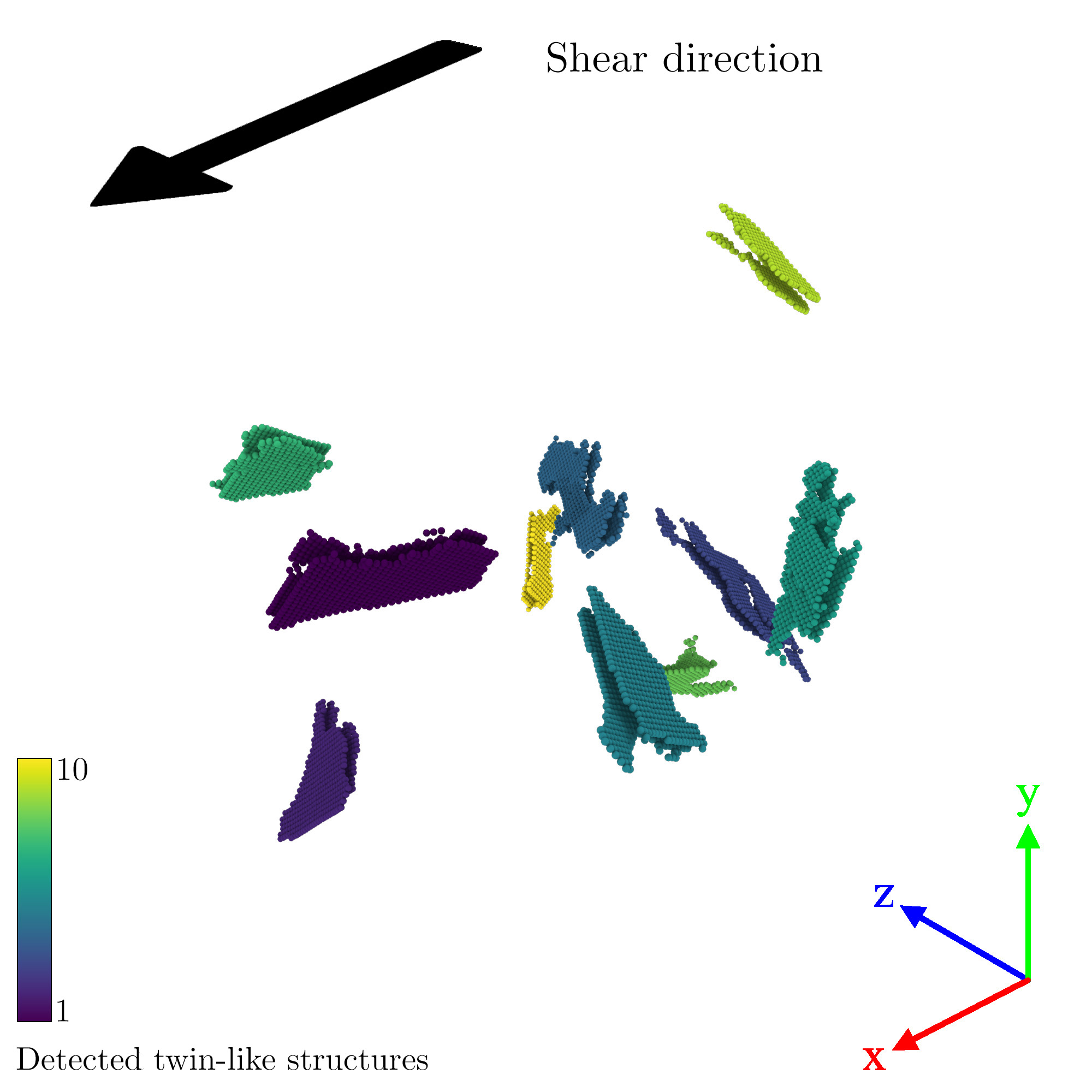}
         \caption{Remaining structures after sorting out with PCA and pairing of parallel planes}
         \label{fig:CSPOsstwins}
     \end{subfigure}
    \caption{Interim results of twin finding algorithm after 28~ps, corresponding to 2.8~nm of shearing.}
    \label{fig:ClusterAndPair}
\end{figure}

The algorithm for twin boundary atom identification~\ref{alg:findTB} results in considerable noise of monatomic and amorphous clusters (see Fig.~\ref{fig:CSCLus}).
As larger structures are expected and of more interest, a generous cluster size threshold of 100 atoms was set to de-clutter the picture.
Non-planar accumulations were removed after being identified via PCA (Alg.~\ref{alg:paramPl}).
Of over 6000 found clusters, only 40 were of the desired shape, and 20 were successfully paired (see Fig.~\ref{fig:CSPOsstwins}). 

\begin{figure}[ht]
    \centering
    \includegraphics[width=0.5\textwidth]{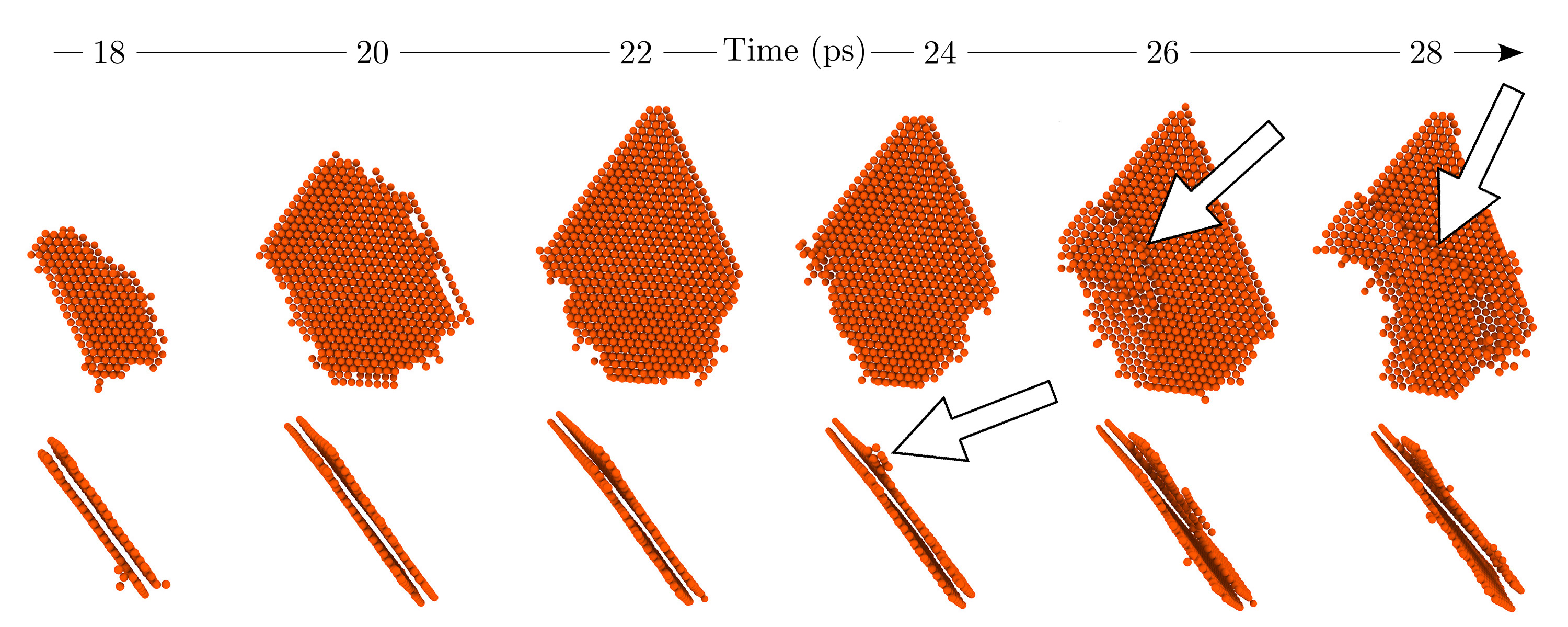}
    \caption{Tracking of single twin, movement of Shockley partials clearly visible.}
    \label{fig:TrackingTwin15}
\end{figure}

Figure~\ref{fig:TrackingTwin15} shows a twin that was successfully tracked over several frames.
At $t = 18-22$~ps, we see the moving Shockley partial dislocation that sets the twin apart from a regular stacking fault, thus contributing to the twin's growth.
Results after 24~ps demonstrate that because of crystal rotation, distortion, and other external influencing factors, the selection of atoms belonging to the twin boundaries cannot be perfect.
Nevertheless, the rough shape remains constant, allowing the twin to be tracked.
The time range $t = 24-28~$ps illustrates the next twin layer building up.

\begin{figure}[b]
    \centering
    \includegraphics[width=0.45\textwidth]{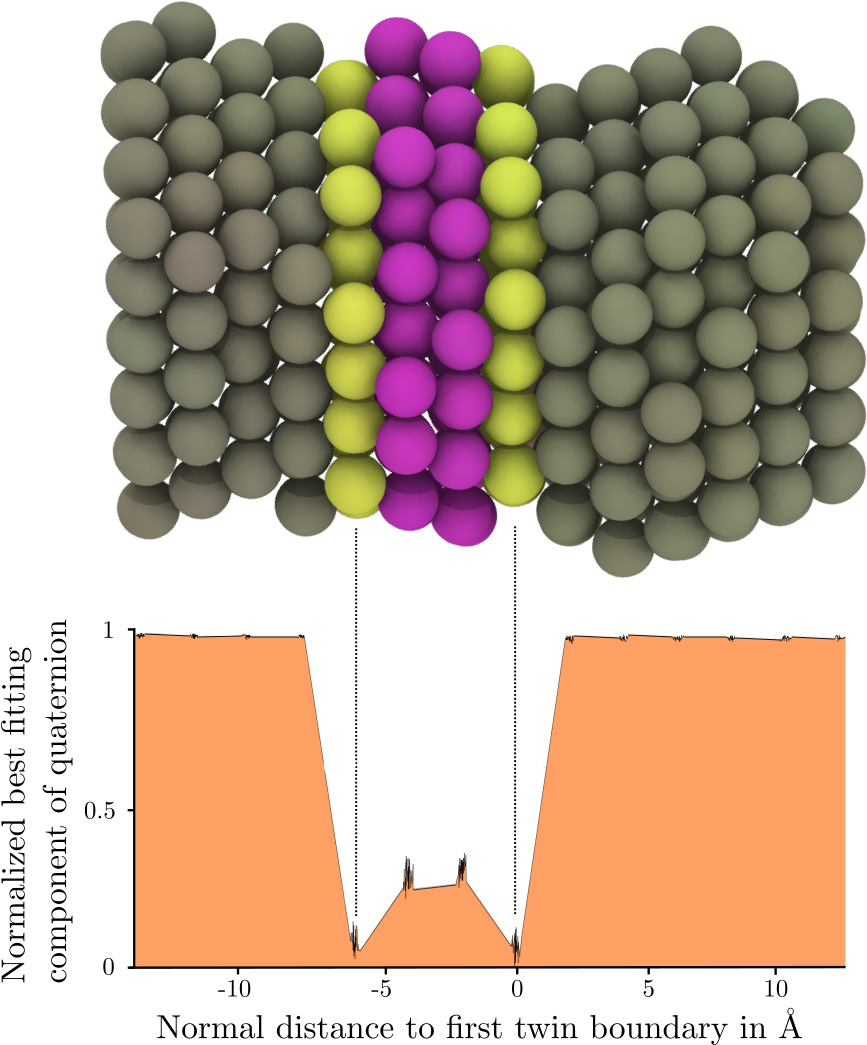}
    \caption{Selection of atoms used to capture and compare the orientation of the outside fcc regions (dark gray) and the twinned volume (violet), separated by the twin boundaries (yellow). The color scheme was derived after translating each per-particle quaternion into an RGB color. The best fitting quaternion-component found in algorithm~\ref{alg:speeradaptive} is plotted against the normal distance to one twin boundary.}
    \label{fig:CSGoodExample}
\end{figure}

Figure~\ref{fig:CSGoodExample} offers a detailed look on a clean and readily identifiable example of a four-layer twin.
No step or moving Shockley partial is captured within the selecion and no disfiguring holes or crystal defects interfere with the twin boundaries.
Thus, the plot of the quaternion component describing the crystallographic orientation, used to automatically identify the separated outside matrix regions from the twinned regions, exhibits a clear shape that allows for a distinction of twinned and regular matrix volume for the selection. 

Calculating the rotation axis and angle twice, once using quaternions from the left region and once from the right region, confirms the twin characteristics. This yields an angle of 59.89° with a 0.28° deviation from the normal vector of the twin boundary plane for the left-side approach, and 60.40° with a 2.67° deviation for the right-side approach, respectively.



\begin{figure}[ht]
    \centering
    \begin{subfigure}[t]{0.35\textwidth}
        \centering
        \includegraphics[width=0.5\textwidth]{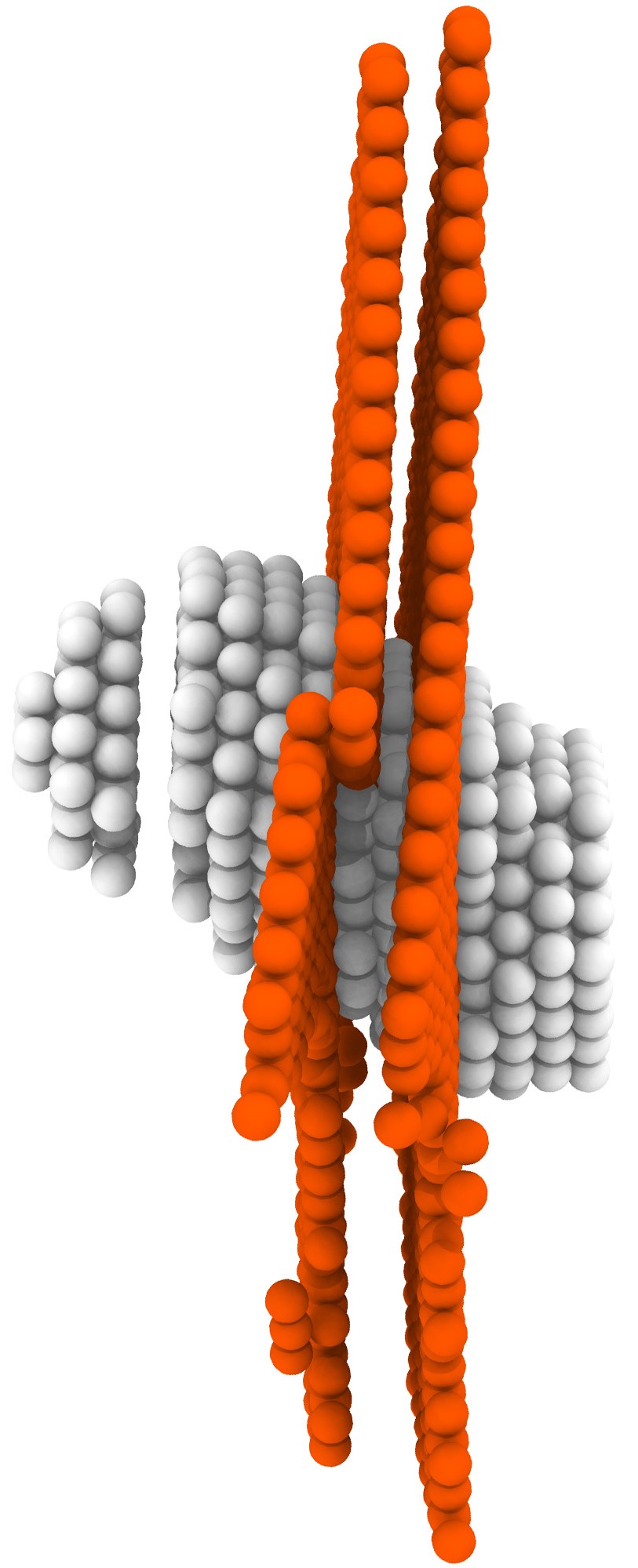}
         \caption{Twin boundaries (orange) and selected atoms for orientation and displacement analysis (white).}
         \label{fig:Twin15}
    \end{subfigure}
     
     \begin{subfigure}[t]{0.45\textwidth}
         \centering
         \includegraphics[width=\textwidth]{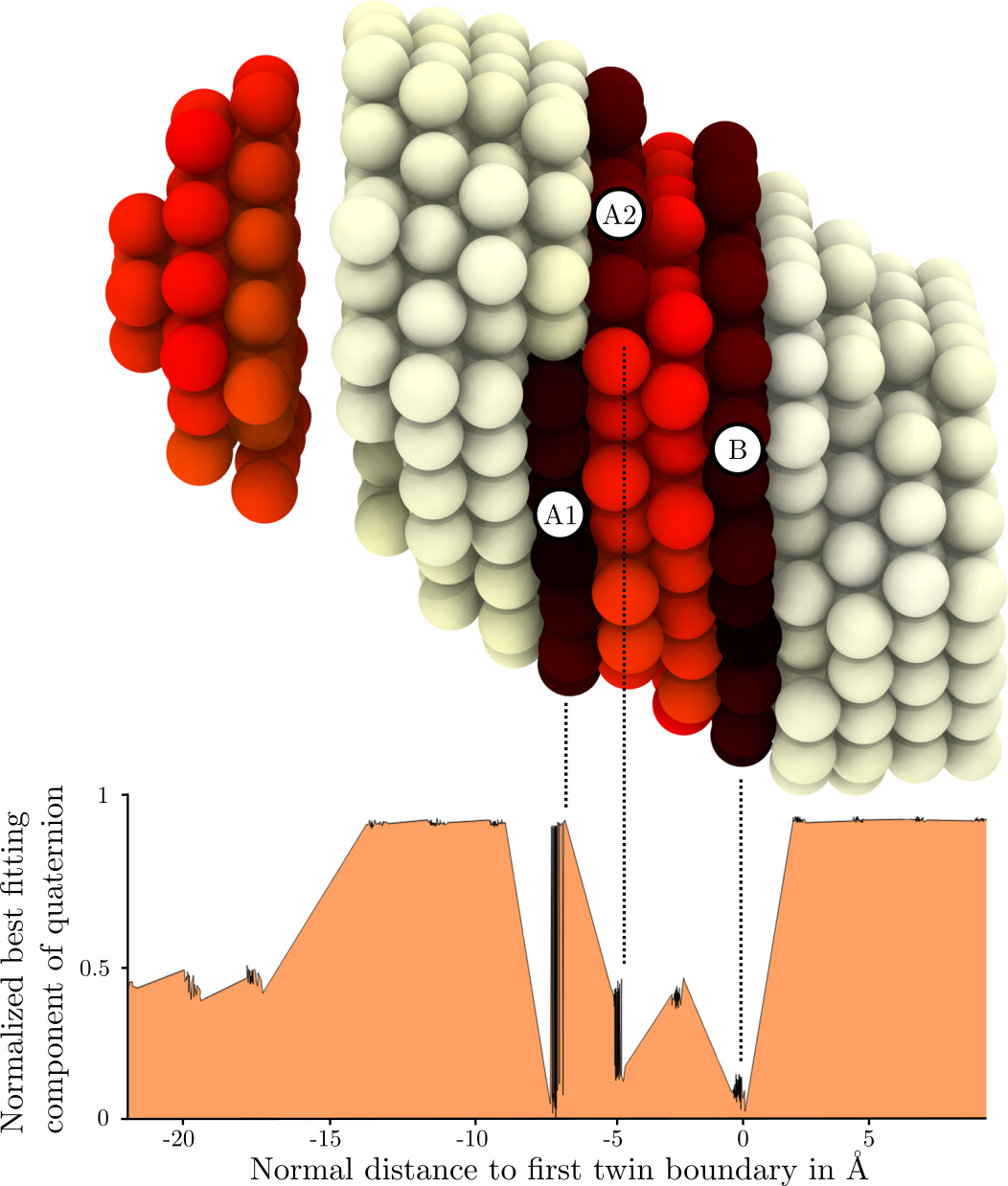}
         \caption{Region selected by algorithm \ref{alg:speeradaptive} for analyzing orientation and displacement and plotted orientation. Twin boundaries are indicated with the letters A and B. Boundary A is subdivided (A1, A2) due to a moving Shockley partial being caught in the selection.}
    \label{fig:CSPoorExample}
     \end{subfigure}
     \caption{Case of twin containing a step that complicates automatic validation.} 
\end{figure}

However, not all twins can be identified so easily.
Further inspecting the same twin introduced in figure~\ref{fig:TrackingTwin15} at $t=28~$ps reveals interesting as well as challenging details.

Figure~\ref{fig:CSPoorExample} demonstrates that, in contrast to the right twin boundary B, its left counterpart is divided into two. 
The segment labeled A2 lies within the twinned region and shares a vertical \{111\} plane with it, which leads to a broader scattering of orientation values along the vertical axis.
Segment A1 forms the remainder of the left twin boundary and shares its layer with some of the outside fcc lattice, exhibiting a significant difference in orientation that results in an even larger spread among the values.
While the reason for the split twin boundary can be a step (a consequence of the deformation twin's shape), figure~\ref{fig:TrackingTwin15} suggests that it corresponds to a forming Shockley partial.
In this case, as the cluster atom furthest from the right twin boundary happens to be a left twin boundary atom, the graph is again characterized by a jump, making rapid analysis by eye difficult.
The gap between the light red and white region belongs to a different planar structure that was excluded from the evaluation in order to avoid jumps in the data.
Despite these challenges, computing the median for each cluster effectively smooths the plot, making automatic validation possible.
Due to the ambiguous position of the left twin boundary, the calculated angle and axis deviation from the left-side approach deviate significantly from the expected values of 60° and 0°, respectively.
The orientation difference is 61.56°, which is within an acceptable range.
However, the axis deviation measures 12.61°.
This underscores the importance of cross-verification, as the right-side approach, considering the planar twin boundary, yields an angle of 59.97° and a deviation between rotation axis and normal vector of only 0.23°.

\section{Challenges and Improvements}
\label{sec:discuss}

\begin{figure}[ht]
     \centering
     \includegraphics[width=0.25\textwidth]{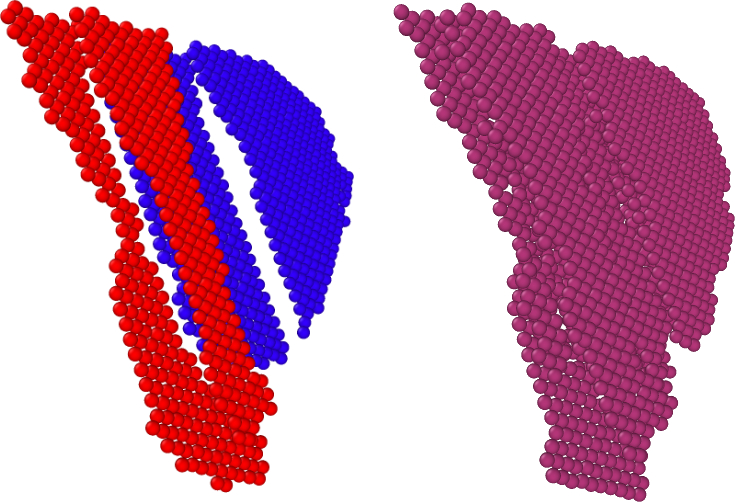}
    \caption{Comparison of results of algorithm~\ref{alg:findTB} before (left) and after (right) revision of the same. The large gap between two formerly individual pairs (red and blue) is closed by including atoms outside a well-defined plane leading to the detection of the entire structure (purple).}
    \label{fig:NewPlanarMethodResults}
\end{figure}

This section discusses the issues and other challenges encountered during twin identification and tracking.
The first attempt to identify planes showed, on closer examination of individual twins, that structures belonging together were not correctly identified as such if the selection criterion was kept too strict.
Finding single layers by the coordination number of hcp-like atoms does not account for possible steps within twin boundaries that form due to the mechanism of deformation twinning.
Studying the structural environment of single layers revealed that atoms of unidentified lattice type could act as the connecting links between well-defined planes.
Adapting algorithm~\ref{alg:findTB} to this knowledge not only dramatically improved the pairing process, but finally allowed to detect twin boundaries in their actual lenticular shape (see Fig.~\ref{fig:NewPlanarMethodResults}).
However, by relaxing the conditions, more atoms are selected, which may establish unwanted connections to other structures.
The identification of linking atoms is a substantial element of the algorithm that is currently being optimized towards more accurately detecting the twin boundary's entire shape on one hand, and reducing the number of artificial connections on the other.

Several of the empirically derived values for the conditions for pairing twin boundaries and tracking twins listed in table~\ref{tab:planeParams} remain debatable.
Unlike indisputable parameters such as average neighbor distance obtained from an RDF-analysis or coordination numbers of atoms in a perfect fcc lattice, the most convenient values for pairing and tracking can slightly change at every timestep.
The pairing algorithm~\ref{alg:findplanes} contains a design decision that may have a noticeable negative impact on the outcome in some cases. 
Whenever multiple parallel matches for one plane are found, the two closest clusters are paired.
Thus, the adjacent twin boundaries of two parallel and close twins might be falsely paired, and since the tracking algorithm is built into the autonomous script, it is possible for non-twins to be erroneously tracked.
This can be resolved by performing the orientation analysis parallel to the pairing process.
First, the conditions must be mitigated so that a broader selection of possible matches is made.
That selection can be filtered by considering the orientation between supposed twin and outside region.
The implementation of this approach would significantly improve the validity of the pairing.
The final pairing would no longer rely on empirically determined values, but on values based on clear references in the literature.
The current separation of automatic pairing and tracking and interactive OVITO session was chosen for a better insight and reduced computational costs during testing.


While the case study focused on showcasing the robust capabilities of our tool in various aspects, it's worth noting that we did not present the results of the displacement analysis introduced in~\ref{sec:interactive}.
At the time of this research, development and testing of this particular feature were still in their initial phases.
Linking the results of the displacement analysis to concrete local stress fields required more extensive testing and validation than was possible within the scope of this study.
However, we emphasize the inherent potential of this analysis.
As a tool in continuous development, it holds promise for future research and applications.
Our future goals include further testing and refinement to provide a meaningful link between displacement and local twin formation stress.

At this point, it should be noted that OVITO has recently introduced an implementation for identifying planar defects in fcc lattices~\cite{identifyplanarOVITO}.
Our tests revealed that the modifier robustly identifies coherent twin boundaries, which is suitable for finding annealing twins.
However, it falls short when it comes to detecting deformation twins, due to their out-of-plane shape. 
By contrast, our tool can detect twin boundaries that extend over multiple planes, offering a more comprehensive analysis.
Most notably, our tool distinguishes itself through the capacity to validate its own findings, employing an orientation analysis between twinned volume and regular fcc region, thereby ensuring a higher degree of accuracy and reliability.


\section{Conclusion and Outlook}

The objective of this work was to develop a post-processing analysis tool for molecular dynamics simulations that allows a comprehensive examination of deformation twins in fcc lattice materials.
In detail, the goals were (1) identification of coherent twin boundaries, (2) connection of boundary pairs, (3) displacement analysis of validated twins and (4) their temporal tracking.
These goals were pursued by (1) implementing an algorithm taking advantage of OVITO's capabilities to detect twin boundaries as hcp-like single layers, (2) plane parameterization by PCA and pairing based on simple geometric relations, (3) orientation and displacement analysis of sensibly selected regions, and (4) cross-timestep spatial analysis.

The results show that twin structures can be detected, tracked, and even examined individually.
We can clearly distinguish between stacking faults and twins, while a temporal analysis allows detailed views and an understanding of the formation mechanism of individual twins.

However, some challenges appearing during the methods development could not yet be solved universally.
The first selection of possibly planar structures is crucial, and inaccuracies at this stage degrade the results of all later computations.
Sub-graph-matching could be a promising alternative for the empirically determined conditions, as successfully proven by Stukowski~\cite{ToolStukowski}.
In the course of this work, it became apparent that most of the difficulties stem from empirically determined numerical values used as hard-coded conditions.

Nonetheless, this tool can already provide valuable insights into twin formation by detecting the majority of twin-like structures and allowing reliable validation of the same.
Especially in sufficiently large systems, where twinning is the preferred deformation mechanism over dislocation slip, good results can be obtained.
Materials scientists may use the tool to collect quantifiable data and examine individual structures, enabling them to better study the formation mechanisms of deformation twins in complex systems.





\section*{Acknowledgement}
Part of this work was funded by the Austrian COMET-Pro\-gram (Project K2 InTribology1, no. 872176) and carried out at the Austrian Excellence Center for Tribology (AC2T research GmbH). 
A.D. acknowledges the German Research Foundation (DFG) under project GR~4174/5-1.
The computational results presented here were obtained using the Vienna Scientific Cluster (VSC).
Open access funding was provided by TU Wien (TUW).

\printcredits

\section*{Declaration of Generative AI and AI-assisted technologies in the writing process}

During the preparation of this work the authors used GPT-3.5 in order to shorten and improve the readability of the introduction. After using this tool, the authors reviewed and edited the section as needed and take full responsibility for the content of the publication.

\section*{Data Availability Statement}
The code used in this study is available on GitHub at \url{https://github.com/hehrich/automated-twin-tracking}.

\FloatBarrier
\bibliographystyle{unsrt}



\end{document}